\documentclass[12pt,a4paper]{article}
\pdfoutput=1

\usepackage{amsmath,amssymb,amsfonts,mathrsfs,mathtools,graphicx,bbm}
\usepackage{fixltx2e}
\usepackage[font=small]{caption}
\usepackage{lmodern}
\usepackage{booktabs}
\usepackage{dsfont}
\usepackage{slashed}
\usepackage{braket}
\usepackage{xspace}

\usepackage{epsfig}
\usepackage{relsize}
\input epsf
\usepackage{enumitem}
\usepackage{color}

\usepackage{hyperref}

\newcommand{\be}{\begin{equation}}
\newcommand{\ee}{\end{equation}}
\newcommand{\bea}{\begin{eqnarray}}
\newcommand{\eea}{\end{eqnarray}}
\newcommand{\bee}{\begin{enumerate}}
\newcommand{\eee}{\end{enumerate}}
\newcommand{\bei}{\begin{itemize}}
\newcommand{\eei}{\end{itemize}}
\newcommand{\bal}{\begin{equation}\begin{aligned}}
\newcommand{\eal}{\end{aligned}\end{equation}}
\newcommand{\bem}{\left (\begin{matrix}}
\newcommand{\eem}{\end{matrix} \right )}

\newcommand{\nn}{\nonumber}
\newcommand{\la}{\label}

\newcommand{\alg}[1]{\mathfrak{#1}}

\def\sl2{\alg{sl}(2)}

\numberwithin{equation}{section}
\makeatletter
 \let\old@startsection=\@startsection
 \let\oldl@section=\l@section
 \renewcommand{\@startsection}[6]{\old@startsection{#1}{#2}{#3}{#4}{#5}{#6\mathversion{bold}}}
 \renewcommand{\l@section}[2]{\oldl@section{\mathversion{bold}#1}{#2}}
\makeatother
\newcommand{\AdS}{\text{AdS}}

\definecolor{grey}{rgb}{0.4,0.4,0.5}
\definecolor{darkgreen}{rgb}{0,0.5,0}
\definecolor{darkred}{rgb}{0.6,0.0,0}
\definecolor{lightbrown}{rgb}{1,0.9,0.8}
\definecolor{brown}{rgb}{0.6,0.3,0.3}
\definecolor{darkblue}{rgb}{0,0,0.5}
\definecolor{darkmagenta}{rgb}{0.5,0,0.5}

\definecolor{TCDblue}{rgb}{0.1,0.3,0.6}

\def\pa {\partial}
\def\ov{\over}
\def\bra{\langle}
\def\ket{\rangle}

\def\a {\alpha}
\def\b {\beta}
\def\g {\gamma}

\def\de {\delta}

\def\e{\epsilon}

\def\vk{\varkappa}

\def\s {\sigma}

\def\z {\zeta}

\def\p{\phi}

\def\T{\Theta}

\def\u\upsilon
\def\U\Upsilon
\def\vU\varUpsilon
\def\vPi{\varPi}

\def\cA{{\cal A}}

\def\cC{{\cal C}}

\def\cE{{\cal E}}

\def\cG{{\cal G}}
\def\cH{{\cal H}}

\def\cJ{{\cal J}}
\def\cK{{\cal K}}
\def\cL{{\cal L}}
\def\cM{{\cal M}}

\def\cO{{\cal O}}
\def\cP{{\cal P}}
\def\cQ{{\cal Q}}
\def\cR{{\cal R}}
\def\cS{{\cal S}}
\def\cT{{\cal T}}
\def\cU{{\cal U}}

\def\cX{{\cal X}}

\def\mS{{\mathscr S}}

\def\bI{{\mathbb I}}
\def\bJ{{\mathbb J}}

\def\bP{{\mathbb P}}

\def\bT{{\mathbb T}}

\def\bZ{{\mathbb Z}}

\def\bfH{{\bf H}}

\def\bfP{{\bf P}}

\def\rJ{{\rm J}}

\def\rQ{{\rm Q}}

\def\th{{\tilde h}}

\def\ha{{\hat a}}
\def\hb{{\hat b}}

\def\x'{\mathaccent 19 x}
\def\y'{\mathaccent 19 y}
\def\n'{\mathaccent 19 n}
\def\u'{\mathaccent 19 u}

\def\et'{\mathaccent 19 \eta}
\def\th'{\mathaccent 19 \theta}
\def\lam'{\mathaccent 19 \lambda}
\def\varet'{\mathaccent 19 \vartheta}
\def\rh'{\mathaccent 19 \rho}
\def\ph'{\mathaccent 19 \phi}
\def\xb'{\mathaccent 19 {\bar{x}}}

\def\d1{{\dot{1}}}

\def\sun2{SU(N)$\times$SU(N)}

\def\Tb{{\overline{T}}}
\def\TTb{${T\overline{T}}$}
\def\JTb{${\rJ\overline{T}}$}
\def\JbT{${\bar{\rJ}T}$}

\def\sUpsilon{\mbox{\scriptsize $\Upsilon$}}

\def\aJTs{\alpha_{\mbox{\tiny ${JT_1}$}}}
\def\aJtTs{\alpha_{\mbox{\tiny ${\widetilde{J}T_1}$}}}
\def\aJTt{\alpha_{\mbox{\tiny ${JT_0}$}}}
\def\aJtTt{\alpha_{\mbox{\tiny ${\widetilde{J}T_0}$}}}

\def\aTTb{\alpha_{\mbox{\tiny ${T\overline{T}}$}}}

\begin{document}

\null\vskip-40pt
 \vskip-5pt \hfill
\vskip-5pt \hfill {\tt\footnotesize
TCDMATH 19-15}
% \vskip-5pt \hfill {\tt\footnotesize HMI-11-05}

\renewcommand{\thefootnote}{\fnsymbol{footnote}}

\vskip 1cm \vskip0.2truecm
\begin{center}
\begin{center}
\vskip 0.8truecm {\Large\bf $T\overline{T}$, $\widetilde J J$, $JT$ and $\widetilde JT$ deformations
}
\end{center}

\vskip 0.9truecm
Sergey Frolov\footnote[2]{ Correspondent fellow at
Steklov Mathematical Institute, Moscow.}\footnote[3]{email: 
frolovs@maths.tcd.ie}
 \\
\vskip 0.5cm

{\it School of Mathematics and Hamilton Mathematics Institute, \\
Trinity College, Dublin 2,
Ireland}

\end{center}
\vskip 1cm \noindent\centerline{\bf Abstract} \vskip 0.2cm 

The light-cone gauge approach to \TTb\ deformed models is generalised to models deformed by U(1) conserved currents $J^\a$, $\widetilde J^\a$, stress-energy tensor $T^\a{}_\b$, and their various quadratic  combinations of the form $\e_{\a\b} K_1^\a K_2^\b$. It is then applied 
to derive a ten-parameter deformed Hamiltonian for a system  of scalars with an arbitrary potential, the flow equations for the Hamiltonian density, and the flow equations for the energy of the deformed model.  
The flow equations disagree with the ones recently proposed 
in arXiv:1903.07606.
The results obtained are applied to analyse a CFT with left- and right-moving conserved currents deformed by these operators. It is shown that with a proper choice of the parameter of the \TTb\ deformation the deformed CFT Hamiltonian density  is independent of the parameters of the $J\T$ and $\bar J\overline \T$ deformations. This leads to the existence of two extra relations which generalise the 
$J\T=0$ and $\bar J\overline \T=0$ relations of the undeformed CFT. The spectrum of the deformed CFT is found and shown to satisfy the flow equations.

\flushbottom

\newpage

\tableofcontents

\renewcommand{\thefootnote}{\arabic{footnote}}
\setcounter{footnote}{0}

%%%%%%%%%%%%%%%%%%%%%%%%%%%%%%%%%%%%%%%%%%
\section{Introduction }
%%%%%%%%%%%%%%%%%%%%%%%%%%%%%%%%%%%%%%%%%%

The \TTb\ deformation of 2d field theories  introduced in \cite{Z04} admits various generalisations if the model under consideration possesses additional  conserved currents. Examples include the Lorentz invariance preserving higher-spin  deformations of integrable models \cite{SZ16}, and the $\rJ\overline{T}$ type deformations \cite{Guica17} which break Lorentz invariance of an undeformed model. Any such a deformation is obtained by adding to the Hamiltonian of an undeformed model an operator of the form 
$\cO_{K_1K_2} \sim \e_{\a\b}K^\a_1K^\b_2$ where $K_1$ and $K_2$ are conserved currents. 
It is hoped that the spectrum of any model deformed by these operators is completely fixed by the spectrum of the undeformed model just as it is for the \TTb\ deformation \cite{Z04}. Indeed, the spectrum of 
a  CFT with left- and right-moving conserved currents
deformed by $\rJ\Tb$ was derived in 
\cite{Kutasov2018}, completing the results of \cite{Guica17}, see also \cite{Guica18}, and  the spectrum of such a CFT deformed by \TTb, $\rJ\Tb$, \JbT\  was conjectured in \cite{Kutasov2019} by relating the problem to deformations of holographically dual strings on $AdS_3$, and it was reproduced in \cite{Hashimoto} from the torus partition sum.  A very general nine-parameter deformation of a CFT  by the stress energy tensor $T$,  the conserved currents $\rJ$ and $\bar \rJ$, and their various quadratic  combinations  was studied in \cite{Mezei2019a}  where a set of flow equations with respect to the deformation parameters was proposed, and then used to conjecture  the spectrum of a deformed CFT.

\medskip

In this paper we generalise the light-cone gauge approach \cite{Sfondrini18b,SF19a} to \TTb\ deformed models 
to the case of the most general ten-parameter deformation
by conserved currents $J^\a$, $\widetilde J^\a$,  the canonical Noether stress-energy tensor $T^\a{}_\b$, and their various quadratic  combinations of the form $\e_{\a\b}K^\a_1K^\b_2$. The light-cone gauge approach was used in \cite{SF19a}
to derive the \TTb\ deformed action for a very general system of any number of scalars, fermions and chiral bosons with an arbitrary potential. 
Basically all of the \TTb\ deformed models \cite{AAF,Tateo16,Bonelli18,Tateo18a,Sfondrini18b,Sfondrini19a,Sethi18,Sfondrini19b,Freedman19a} studied before and after \cite{SF19a} are  particular cases of this system  which  includes various Lorentz invariant systems of bosons and fermions, in particular  supersymmetric sigma models, and some of non-Lorentz invariant systems, e.g. the chiral SYK model and the nonlinear matrix 
Schr\"odinger model.

\medskip

The light-cone gauge approach is based on the observation that 
the homogeneous inviscid Burgers equation which determines the spectrum of a \TTb\ deformed model with zero momentum coincides with the gauge invariance condition of the target space-time energy and momentum of a non-critical string theory quantised in a parameter dependent light-cone gauge introduced in \cite{AFZ06a}.
The light-cone gauge-fixed Hamiltonian can be thought of as the Hamiltonian of a deformed model, and the deformation parameter can be identified with the light-cone gauge parameter, see \cite{SF19a} for a detailed discussion.  
The deformed Hamiltonian $\cH_\a$ can be used to derive the flow equation,
$\pa_\a\cH_\a=\,$\TTb, with respect to the  parameter $\a$ of the \TTb\ deformation. 
The flow equation is then used to get  the  inviscid Burgers equation which governs the spectrum of the deformed model. 

\medskip

In this paper we only consider a system of $n$ scalars with an arbitrary potential.
We assume that the bosonic model is invariant under shifts of one of its fields, say $x^1$, and $J^\a$ is its canonical Noether current due to the symmetry. The current 
$\widetilde J^\a$ is a topological current associated to the field $x^1$: $\widetilde J^\a=\e^{\a\b}\pa_\b x^1$. In the case of a CFT the left- and right-moving conserved currents are linear combinations of $J^\a$ and $\widetilde J^\a$. 
We begin with the usual action for bosonic strings invariant under shifts of three  isometry coordinates $x^a$,  $a=+,-,1$. The undeformed model is obtained by imposing the standard light-cone gauge $x^+=\tau\,,\, p_-=1$ where $p_-$ is the momentum conjugate to $x^-$.  
It appears that to describe the most general  ten-parameter deformation  we need to introduce pairs of auxiliary non-dynamical co-vectors and scalars $(V^a_\a\,,\, \tilde X_a)$ and $(\cU_{a\a}\,,\, \Upsilon^a)$ associated with the conserved currents $J_a^\a$ and $\widetilde J^{a\a}$, respectively. We then perform a ten-parameter canonical transformation $A$ which involves the three coordinates $x^a$ and their momenta $p_a$, and the coordinates and momenta of the auxiliary fields. Finally, we impose the light-cone gauge $x^+=\tau\,,\, p_-=1$ on the transformed coordinates and momenta, solve the Virasoro and Gauss-law constraints, and identify the light-cone gauge-fixed Hamiltonian $\cH_A=-p_+$ with the deformed one.

\medskip

The derivatives of the deformed Hamiltonian with respect to the ten  parameters
give rise to flow equations (\ref{eqvp}-\ref{eqa1m}) which are valid for any bosonic model of the type we study. We believe that the flow equations are universal, and apply
to any Lorentz and non-Lorentz invariant model of bosons and fermions.  
The flow equations disagree with  the ones recently proposed in \cite{Mezei2019a}.
An obvious difference is that some of the coefficients in front of the deforming operators on the r.h.s. of our flow equations depend on the deformation parameters of the quadratic operators while those of the flow equations of \cite{Mezei2019a} do not, see section 3 for a detailed discussion.

\medskip

The flow equations for the deformed Hamiltonian can be used to find a system of    flow equations (\ref{eqa11en}, \ref{eqapmen}) for the spectrum of the deformed model. The system does not involve derivatives of the energy with respect to the parameters of the linear deformations by the space components of the currents $J^\a$ and $\tilde J^\a$. These derivatives were used in \cite{Mezei2019a} to find the expectation values of $J^1$ and $\tilde J^1$. As we explain in the paper, these expectation values are given by the derivatives of the energy with respect to the charges $\widetilde P^1$ and $P_1$ of the currents $J^\a$ and $\tilde J^\a$. Thus, we do not really need the two parameters.  The system of flow equations is complicated and at the moment we do not know how to find a proper solution of the system for a generic state. Nevertheless, we show that if a state satisfies level-matching conditions which we derive then the energy of such a state obeys a version of the
homogeneous inviscid Burgers equation.

\medskip

The flow equations for the energy however can be solved for a deformed CFT with left- and right-moving conserved currents. The solution  (\ref{eqEsol})  we find reduces to the known ones in the particular cases of one- and two-parameter deformations. In the case of the \TTb, $\rJ\Tb$, \JbT\ deformation  our solution is equivalent to the one proposed in \cite{Kutasov2019,Hashimoto}   after a proper  redefinition of the deformation parameters, see \eqref{compHK}. It seems to agree with the one proposed in \cite{Mezei2019a}
at least if one switches off their parameters of the  deformation by $\rQ_\pm^1$, and parameters of the $\rJ \T$ and $\bar\rJ \overline{\T}$ deformations.
To get a precise agreement it might be necessary to perform an extra redefinition of the parameter of the \TTb\ deformation. We have not tried to do it.

\medskip

The plan of the paper is as follows. In section 2.1 we introduce an extended action for bosonic strings propagating in a target manifold possessing (at least) three
abelian isometries and determine all the constraints and gauge conditions necessary to recover the undeformed model. In section 2.2 we generate a ten-parameter deformation of the model as a chain of one-parameter canonical transformations producing different deformations. In section 3.1 we impose the light-cone gauge on the transformed coordinates and momenta, solve all the constraints, and find the deformed Hamiltonian. In section 3.2 we use the Hamiltonian to derive the flow equations with respect to the deformation parameters. Then in section 3.3 the flow equations are converted into the flow equations for the spectrum of the deformed model.
 In section  3.4 we determine level-matching conditions physical states must satisfy, and derive the
homogeneous inviscid Burgers equations for the energy of those states.
The flow equations for the Hamiltonian density and the spectrum are then rewritten in the CFT conventions in section 4 where we also consider a very special case of a CFT with left- and right-moving conserved currents, and show that for such a CFT there are  extra constraints \eqref{eqbep3c} replacing the relations 
$\rJ \T=0$ and $\bar\rJ \overline{\T}=0$ of the undeformed CFT. In section 4.3 we propose a formula for the spectrum of such a deformed CFT which satisfies the flow equations.
In Conclusions we discuss open questions and generalisations of the light-cone gauge approach. Finally, in Appendices we collect some explicit formulae, and consider examples of deformed Hamiltonians, and deformed spectrum of CFT with left- and right-moving conserved currents.

%%%%%%%%%%%%%%%%%%%%%%%%%%%%%%%%%%%%%%%%%%
\section{String model and the transformation}\la{blc}
%%%%%%%%%%%%%%%%%%%%%%%%%%%%%%%%%%%%%%%%%%
\subsection{Extended string model}\la{extstrs}

In this section we follow the same approach as in \cite{SF19a} but simplify a bit their consideration. We begin with bosonic strings propagating in a $n+2-$dimensional target Minkowski manifold $\cM$ possessing
(at least) three abelian isometries realised by shifts of the  isometry
coordinates.  
We denote coordinates of $\cM$ by $x^M$, $M=+,-,1,\ldots,n$,
and  choose the  isometry
coordinates  to be\footnote{$x^\pm$ are related to the  time and space isometry coordinates $t$ and $\p$ used in \cite{SF19a} as follows: 

$x^- =\p \,-\,t\ , \,\,
x^+ ={1\ov2}(\p\, +\,t)$.} $\, x^+ $,  $x^{-}$ and $x^1$. 
The  ``transversal'' coordinates  $x^\mu$, $\mu=1,\ldots,n$ are the fields of the model we wish to deform.
 Obviously, the target-space metric $G_{MN}$ of $\cM$ does not depend on  $x^\pm $ and $x^1$. 
 
 \medskip

Assuming for simplicity that the $B$-field vanishes but making no assumption on the form of $G_{MN}$, we write
the initial string action in the standard form
\bal \la{S1} 
S = \int_{0}^{R}\, {\rm
d}\s{\rm d}\tau\, \cL\,,\quad \cL=-{1\over 2}\g^{\a\b}\partial_\a x^M \partial_\b x^NG_{MN} 
\,, 
\eal 
where $\gamma^{\a\b}= h^{\a\b} \sqrt {-h}$  is the Weyl-invariant
combination of the world-sheet
metric $h_{\a\beta}$ with $\det\gamma=-1$, and $\e^{01}=-\e_{01}=1$. The range $R$ of the world-sheet space coordinate $\s$  will be fixed by a generalised uniform light-cone gauge.

The  string action invariance under the shifts of $x^\pm $ and $x^1$
leads to the existence of the three conserved currents
\bal\nn
J_a^\a = {\pa\cL\ov \pa \pa_\a x^a}\,,\quad \pa_\a J_a^\a =0\,,\quad a=+,-,1\,.
\eal
Then, to any coordinate $x^M$ we can associate a topological conserved current 
\bal\nn
\widetilde J^{M\a}=\e^{\a\b}\pa_\b x^M\,,
\eal
and the deformations we discuss in this paper involve the  topological current $\tilde J^{\a}\equiv\tilde J^{1\a}$
which for a free massless scalar field is dual to $J^\a\equiv J_1^\a$. We denote the corresponding conserved charges as 
\bal\nn
P_a =\int_{0}^{R} {\rm d}\s J_a^0\,,\quad \widetilde P^a =\int_{0}^{R} {\rm d}\s \widetilde J^{a0}\,.
\eal
It is clear that $\widetilde P^a$ does not vanish only if $x^a$ has a nonzero winding number.

\medskip

To get the most general deformation we introduce pairs of non-dynamical co-vectors and scalars $(V^a_\a\,,\, \tilde X_a)$ and $(\cU_{a\a}\,,\, \Upsilon^a)$ associated with  currents $J_a^\a$ and $\tilde J^{a\a}$, respectively. Then, the extended string action together with the new fields has the form
\bal \la{S1b} 
S = \int_{0}^{R}\, {\rm
d}\s{\rm d}\tau\, \Big( -{1\over 2}\g^{\a\b}\partial_\a x^M \partial_\b x^NG_{MN} -\e^{\a\b}V_\a^a\pa_\b \tilde X_a -\e^{\a\b}\cU_{a\a}\pa_\b \Upsilon^a \Big)
\,.
\eal 
The action \eqref{S1} is obviously invariant under reparametrisations,  the six U(1) gauge symmetries
\bal\la{gaugeu1}
V_\a^a\to V_\a^a + \pa_\a\xi^a\,,\quad \tilde X_a\to \tilde X_a\,,\qquad \cU_{a\a}\to \cU_{a\a} + \pa_\a\z_a\,,\quad \Upsilon^a\to \Upsilon^a\,,
\eal
and the constant shifts of the coordinates  $\tilde X_a$ and $\Upsilon^a$
\bal\nn
 \tilde X_a\to \tilde X_a + c_a\,,\quad \Upsilon^a\to\Upsilon^a +\eta^a\,.
\eal
The last symmetry leads to the existence of six conserved currents 
\bal\nn
\pa_\a(\e^{\a\b} V_\b^a)=0\,, \quad \pa_\a(\e^{\a\b} \cU_{a\b})=0\,.
\eal
These are equations of motion for $V_\a^a$ and $\cU_{a\a}$ which imply that
\bal\nn
V_\a^a = \delta_{1\a}v^a +\pa_\a\xi^a  \,,\quad \cU_{a\a} =  \delta_{1\a}u_a +\pa_\a\z_a\,,
\eal
where $v^a$ and $u_a$ are time-independent zero modes of $V_1^a$ and $\cU_{a1}$, respectively. Thus, one gets a family of models parametrised by $v^a$ and $u_a$, and if they vanish then we get back to the original string model \eqref{S1}. We are going to impose a gauge condition which depends on  ten parameters, and consider the resulting gauge-fixed action as a deformation of our favourite one. 

\medskip

The simplest way to impose such a gauge condition is to switch  to the Hamiltonian formalism. Introducing the momenta canonically-conjugate to the coordinates $x^M$
 \bal\nn
 p_M ={\de S\ov \de \dot{x}^M} = -\g^{0\b}\partial_\b x^N\,
G_{MN}\,,\quad \dot{x}^M\equiv \pa_0 x^M \,, 
\eal 
 we bring the
string action (\ref{S1b}) to the first-order form 
\bal \la{S2} 
S=\int_{0}^{R}\, {\rm d}\s{\rm d}\tau\, \left( p_a \dot{x}^a +V^a\dot{\tilde X}_a+\cU_a\dot\Upsilon^a+p_k \dot{x}^k +
{\g^{01}\ov\g^{00}} C_1+ {1\ov  \g^{00}}C_2-V_0^aC_a^V-\cU_{a0}C^a_\cU\right)\,.
\eal
Here $a=+,-,1$, $k=2,\ldots,n$, and $V^a\equiv V^a_1$, $\,\cU_a\equiv \cU_{a1}$. Then,
\bal\nn
C_1&=p_Mx'^M = p_a x'^a+p_k x'^k\,,\quad x'^M\equiv
\pa_1 x^M\,.
\\
C_2&={1\ov2}G^{ab} p_a p_b + {1\ov2}x'^a x'^b G_{ab}+G^{ak} p_a p_k + x'^a x'^k G_{ak}+\widetilde\cH_x\,,
\eal
are the Virasoro constraints, 
and $\widetilde\cH_x$ 
depends only on the transversal fields $x^k$ and $p_k$
\bal\la{Hx} 
\widetilde\cH_x = {1\ov2}G^{kl}p_k p_l +  {1\ov2}x'^k x'^l\,G_{kl}\,,\quad k,l=2,\ldots,n\,.
\eal 
Finally, 
\bal\nn
C_a^V = \tilde X_a' \,,\quad C^a_\cU= \Upsilon'^a\,,\quad a=+,-,1\,,
\eal
are the constraints which generate the U(1) gauge transformations \eqref{gaugeu1}. It is clear from \eqref{S2} that $V^a$ and $\cU_a$ are the momenta canonically-conjugate  to the coordinates $\tilde X_a$ and $\Upsilon^a$.  It is worthwhile to mention that on the constraints surface one can replace 
\bal\la{tC1}
C_1\to \tilde C_1 = p_a x'^a+V^a \tilde X_a'+\cU_a\Upsilon'^a+p_k x'^k\,.
\eal 
This simple observation will be useful later.

\medskip

If we do nothing with the coordinates and momenta of the extended model, and just impose the light-cone gauge
conditions 
\bal \la{ulc} 
x^+ = \tau \,,\quad p_- = 1\,,\quad V'^a=0 \,,\quad \cU'_a=0\,,
\eal
then, solving the constraints $C_a^V =0$ and $C^a_\cU=0$, we get that $V^a=v^a$, $\tilde X_a=\tilde x_a$, $\cU_a=u_a$ and $\Upsilon^a=\sUpsilon^a$ are $\s$-independent.
Using the Virasoro
constraint $C_1$ to find $x'^-$ 
\bal\la{C1c}
  x'^-=- p_\mu x'^\mu\equiv -px'\,,
\eal
we bring $C_2$ to the form
\bal \la{c2b}
C_2 &= {G^{++}\ov2} p_+^2  + (G^{-+}+G^{\mu+}p_\mu)p_+  +{G^{--}\ov2}+ G^{\mu-}p_\mu + {G_{--}\ov2}(px')^2  - G_{\mu-}x'^\mu px' +\cH_x\,, \\
\cH_x &= {1\ov2}G^{\mu\nu}p_\mu p_\nu +  {1\ov2}x'^\mu x'^\nu\,G_{\mu\nu}\,.
\eal 
Solving the resulting equation $C_2=0$  for $p_+$,
 we find the  gauge-fixed action of the extended model
\bal \la{S0} 
S_0=\int{\rm d}\tau\, \left(v^a\dot{\tilde x}_a+u_a\dot{\sUpsilon}^a\right)+\int_{0}^{R}\, {\rm d}\s{\rm d}\tau\, \left( p_\mu \dot{x}^\mu
\,-\, \cH_{0} \right)\, ,\quad \cH_0 \,=\, -p_+(p_\mu, x^\mu ,x'^\mu ) \,.
 \eal
 Here the integration range $R=P_-$ is found by integrating the gauge condition $p_-=1$ over $\s$, and
the density $\cH_0$ of the world-sheet Hamiltonian depends on the periodic
transversal fields. Thus, the gauge-fixed string
action describes a two-dimensional model on a cylinder of
circumference $ R=P_-$. Obviously, $v^a$ and $u_a$ are time-independent, and   setting them to any constants, one gets the usual light-cone gauge string action.   
The gauge-fixed model in general is not Lorentz invariant, and, as was shown in \cite{SF19a}, to get a Lorentz invariant model one has to choose the target space metric of the form
\bal\la{GMNrel}
ds^2 = dx^+dx^- -2V dx^+dx^+ + G_{\mu\nu}dx^\mu dx^\nu\,,
\eal
where $V$ is an arbitrary function of the transversal coordinates. Then,  $\cH_0$ becomes the Hamiltonian density of a sigma-model of $n$ scalar fields with the
potential $V$
\bal\la{ch0}
\cH_0=\cH_x + V(x)\,.
\eal

We are going to perform a parameter dependent canonical transformation of the coordinates $x^a$, $\tilde X_a$, $\Upsilon^a$ and momenta $p_a$, $V^a$, $\cU_a$, and impose the light-cone gauge conditions \eqref{ulc}  on the transformed $x^+$, $p_-$, $V^a$ and $\cU_a$.\footnote{It is worthwhile to mention that in terms of the original coordinates this is a multi-parameter uniform light-cone gauge generalising the one introduced in \cite{AFZ06a}. Technically, it is easier to work with the transformed coordinates and momenta in terms of which the gauge is the standard light-cone one. }
Then, solving the constraints, one gets a parameter-dependent gauge-fixed Hamiltonian density $\cH_A$ where $A$ labels the parameters.
We want to think about this gauge-fixed model as a deformation of the model with the Hamiltonian density $\cH_0$, and consider the parameters of the canonical transformation as the deformation parameters.

\medskip

The deformed two-dimensional model with the Hamiltonian density $\cH_A$ is  invariant under the shifts of
the world-sheet coordinates $\tau$ and $\s$, and its canonical stress-energy tensor is given by
\bal\la{Tmunu}
T^0{}_0&= \cH_A\,,\quad T^1{}_0= -{\pa \cH_A\ov \pa x'^\mu}{\pa \cH_A\ov \pa p_\mu}\,,
\\
T^0{}_1&= p_\mu x'^\mu\,,\quad T^1{}_1=  \cH_A-{\pa \cH_A\ov \pa x'^\mu} x'^\mu   -p_\mu {\pa \cH_A\ov \pa p_\mu}\,.
\eal
In particular,  the 
world-sheet energy and momentum  are conserved
\bal\nn
E = \int_{0}^{R}\, {\rm d}\s\, \cH_A = -P_+\,,\quad  
P = -\int_{0}^{R}\, {\rm d}\s\, p_\mu x'^\mu\,. 
\eal 
Then, the Hamiltonian density is invariant under a shift of the coordinate $x^1$, and the corresponding conserved current is
\bal\la{Jmu}
J^0=p_1\,,\quad J^1= -{\pa \cH_A\ov \pa x'^1}\,.
\eal
Finally, the topological current in the Hamiltonian formalism is given by
\bal\la{Jtmu}
\widetilde J^0=x'^1\,,\quad \widetilde J^1= -\dot x^1=-{\pa \cH_A\ov \pa p_1}\,.
\eal

%%%%%%%%%%%%%%%%%%%%%%%%%%%%%%%%%%%%%%%%%%
\subsection{Parameter dependent canonical transformation}\la{cantransform}

The simplest way to find a canonical transformation generating  a multi-parameter deformation of $\cH_0$ is to realise it as a chain of one-parameter  transformations producing different deformations. 
Analysing infinitesimal transformations, one can identify suitable one-parameter ones. We find convenient to use the following consecutive ones

\medskip

1. Transformation, $\cT_{T\bar T}$, generating the \TTb\ deformation \cite{SF19a}\footnote{The parameter $a_{+-}$ is related to the parameter $a$ of the uniform light-cone gauge \cite{AFZ06a}, and the parameter $\a$ of the \TTb\ deformation \cite{Z04} as follows $a_{+-}=\a=a-{1\ov2}$.}
\bal\nn
\cT_{T\overline T}:\quad x^+\to x^+ - a_{+-}x^-\,,\quad x^-\to x^- \,,\quad p_+\to p_+\,,\quad p_-\to p_- + p_+a_{+-}\,.
\eal
It is easy to check\footnote{See appendix \ref{flow-eq} for a short explanation of how it can be done.} that for any target space metric $G_{MN}$ the gauge-fixed Hamiltonian density $\cH_{a_{+-}}$ satisfies the flow equation
\bal\nn
{\pa\cH_{a_{+-}}\ov\pa a_{+-}} = {T\overline T}\,,\quad {T\overline T}\equiv T^0{}_1T^1{}_0 -T^1{}_1T^0{}_0 = -\e_{\a\b}T^\a{}_1T^\b{}_0=T_1T_0\,.
\eal

2. Transformation, $\cT_{\tilde JT_0}$, generating the $-\widetilde JT_0$ deformation 
\bal\nn
\cT_{\tilde JT_0}:\quad x^+\to x^+ - a_{+1}x^1\,,\quad x^1\to x^1 \,,\quad p_+\to p_+\,,\quad p_1\to p_1 + p_+a_{+1}\,,
\eal
\bal\nn
{\pa\cH_{a_{+1}}\ov\pa a_{+1}} = -\widetilde JT_0\,,\quad \widetilde JT_0\equiv \widetilde J^0T^1{}_0 -\widetilde J^1T^0{}_0 = -\e_{\a\b}\widetilde J^\a T^\b{}_0\,.
\eal

3. Transformation, $\cT_{ JT_1}$, generating the $ JT_1$ deformation 
\bal\nn
\cT_{ JT_1}:\quad x^1\to x^1 - a_{1-}x^-\,,\quad x^-\to x^- \,,\quad p_1\to p_1\,,\quad p_-\to p_- + p_1a_{1-}\,,
\eal
\bal\nn
{\pa\cH_{a_{1-}}\ov\pa a_{1-}} =  JT_1\,,\quad  JT_1\equiv  J^0T^1{}_1 - J^1T^0{}_1 = -\e_{\a\b} J^\a T^\b{}_1\,.
\eal

The  sequence $\cT_{ JT_1}\cT_{\tilde JT_0}\cT_{T\overline T}$ of these three transformations can be represented by a single canonical transformation  
\bal\nn
x^a\to (A^{-1})^a{}_b\,x^b\,,\quad p_a\to p_b\,A^b{}_a\,,\quad a,b=+,-,1\,.
\eal
Here $A$ is the following matrix
\bal\nn
A=\left(
\begin{array}{ccc}
1 & a_{+-} &  a_{+1} \\
  0 &1 & 0 \\
0 &  a_{1-} &1 \\
\end{array}
\right)\,,\ A^{-1}=\left(
\begin{array}{ccc}
 1 &  -\tilde a_{+-}& {- a_{+1}}  \\
   & {1} & 0 \\
  0 &- a_{1-}   & 1  \\
\end{array}
\right)\,,
\eal
where 
\bal\nn
\tilde a_{+-} =a_{+-}-a_{+1} a_{1-}\,.
\eal
In general $A$  can be any nondegenerate real matrix with entries $A_{ij}=\delta_{ij}+a_{ij}$, and  one can easily show that 

(i) $A_{++}$  deforms by $-T^0{}_0$:
 $A_{++}\pa_{A_{++}} \cH_{A_{++}} = - T^0{}_0 = -\cH_{A_{++}}$; 
 
(ii) $A_{--}$ deforms by $T^1{}_1$:
 $A_{--}\pa_{A_{--}} \cH_{A_{--}} =  T^1{}_1$; 
 
(iii) $A_{11}$ deforms by $\widetilde JJ$:
 $A_{11}{\pa\cH_{A_{11}}\ov\pa A_{11}} = {\widetilde JJ}$; 
 
(iv) $a_{-+}$ deforms by $1$:
 $\pa_{a_{-+}} \cH_{a_{-+}} =  1$; 

(v) $a_{-1}$ deforms by $-\widetilde J^1$:
 $\pa_{a_{-1}} \cH_{a_{-1}} =  -\widetilde J^1$; 

(vi) $a_{1+}$ deforms by $ J^0=p_1$:
 $\pa_{a_{1+}} \cH_{a_{1+}} =  J^0$.

\noindent We do not need the deformations by $T^0{}_\a$, $J^0$ and $\widetilde J^0$  because they just change the energy by the corresponding charges. 
We could have used the transformations with $A_{--}$ and $A_{11}$ at the next steps but this leads to complicated flow equations when all ten parameters are nonvanishing, and a careful analysis shows that it is better to use these transformations (which commute) at the very end. As to the transformation with  $a_{-1}$, it appears to be easier to use the auxiliary non-dynamical fields to generate the deformation  by $\widetilde J^1$ and the remaining four deformations.  

\medskip

\noindent The remaining transformations are

\medskip

4. Transformation, $\cT_{ T^1{}_0}$, generating the $ -T^1{}_0$ deformation 
\bal\nn
\cT_{ T^1{}_0}:\quad x'^+\to x'^+ - V^+\,,\quad \tilde X'_+\to \tilde X'_+ + p_+\,,\quad p_+\to p_+\,,\quad V^+\to V^+\,,
\eal
\bal\nn
{\pa\cH_{v^+}\ov\pa v^+} = -T^1{}_0\,.
\eal
This transformation obviously represents gauging the shift symmetry of $x^+$, and due to the light-cone gauge condition $x^+=\tau$ it is natural that the zero mode of $V^+$ generates the deformation by $- T^1{}_0$.

5. Transformation, $\cT_{ J^1}$, generating the $ J^1$ deformation 
\bal\nn
\cT_{ J^1}:\quad x'^1\to x'^1 - V^1\,,\quad \tilde X'_1\to \tilde X'_1 + p_1\,,\quad p_1\to p_1\,,\quad V^1\to V^1\,,
\eal
\bal\nn
{\pa\cH_{v^1}\ov\pa v^1} = J^1\,.
\eal
Gauging the shift symmetry of $x^1$  leads to the dependence of  $\cH_{v^1}$ on $x'^1-v^1$.

6. Transformation, $\cT_{ JT_0}$, generating the $ JT_0$ deformation 
\bal\nn
\cT_{ JT_0}:\quad x^+\to x^+ - b_{+1}\tilde X_1\,,\quad \tilde X_1\to \tilde X_1 \,,\quad p_+\to p_+\,,\quad V^1\to V^1 + p_+b_{+1}\,,
\eal
\bal\nn
{\pa\cH_{b_{+1}} \ov\pa b_{+1}} =  JT_0\,,\quad  JT_0\equiv  J^0T^1{}_0 - J^1T^0{}_0 = -\e_{\a\b} J^\a T^\b{}_0\,.
\eal
This transformation must be performed after the transformation $\cT_{ J^1}$. 

7. Transformation, $\cT_{ \tilde J^1}$, generating the $\widetilde J^1$ deformation 
\bal\nn
\cT_{ \widetilde J^1}:\quad x^1\to x^1\,,\quad \Upsilon^1\to  \Upsilon^1+x^1\,,\quad p_1\to p_1-\cU_1\,,\quad \cU_1\to \cU_1\,,
\eal
\bal\nn
{\pa\cH_{u_1}\ov\pa u_1} = \widetilde J^1\,.
\eal
This transformation in fact represents gauging the symmetry generated by 
\bal\nn
\widetilde Q(\xi)=\int d\s \widetilde J^0 \xi =\int d\s x'^1 \xi  \,,
\eal 
which is trivial unless $\xi$ depends on $\s$.
This leads to the dependence of  $\cH_{u_1}$ on $p_1-u_1$.

8. Transformation, $\cT_{\widetilde JT_1}$, generating the $\widetilde JT_1$ deformation 
\bal\nn
\cT_{\widetilde JT_1}:\quad &\tilde X_-\to \tilde X_- +p_-- {b_{--}\ov c_{1-}}(x^--V^-)+b_{--}(x'^1-V^1)\,,\quad \tilde X_1\to \tilde X_1 + c_{1-} \tilde X_- \,,\quad\\
&x'^-\to x'^--V^- -c_{1-}(x'^1-V^1)\,,\\
& p_-\to p_-+V^-{b_{--}\ov c_{1-}}+b_{--}(x'^1-V^1)\,,\quad p_1\to p_1 - c_{1-} \tilde X'_- -c_{1-}\,,\\
&V^-\to V^- +c_{1-}(x'^1-V^1)\,,
\eal
\bal\nn
{\pa\cH_{b_{--},c_{1-}}\ov\pa b_{--}} =\widetilde JT_1\,,\quad {\pa\cH_{b_{--},c_{1-}}\ov\pa c_{1-} } = 0\,,\quad \widetilde JT_1\equiv \widetilde J^0T^1{}_1 -\widetilde J^1T^0{}_1 = -\e_{\a\b}\widetilde J^\a T^\b{}_1\,,
\eal
where in the gauge-fixed Hamiltonian we set $v^-=0$.
Note that this transformation requires both $b_{--}$ and  $c_{1-}$ to be nonvanishing. This is the most complicated transformation. It can be represented as a combination of three simpler transformations
\bal\nn
\cT_{\tilde JT_1} = \cT_{3}\cT_{2}\cT_{1}\,,
\eal
where $\cT_{1}$ gauges the shift symmetry of $x^-$
\bal\nn
\cT_{1}:\quad x'^-\to x'^- - V^-\,,\quad \tilde X'_-\to \tilde X'_- + p_-\,,\quad p_-\to p_-\,,\quad V^-\to V^-\,,
\eal
$\cT_{2}$ is a  linear transformation
\bal\nn
\cT_{2}:\quad \tilde X'_-\to \tilde X'_-  - {b_{--}\ov c_{1-}}x^-\,,\quad p_-\to p_-+V^-{b_{--}\ov c_{1-}}\,,
\eal
and $\cT_{3}$ is a transformation we call twisting $V^-$ with $x'^1$
\bal\nn
\cT_{3}:\quad & \tilde X_1\to \tilde X_1 + c_{1-} \tilde X_- \,,\quad p_1\to p_1 - c_{1-} \tilde X'_- -c_{1-}\,,\quad V^-\to V^- +c_{1-}(x'^1-V^1)\,.
\eal
We could not find a simpler way to generate the $\widetilde JT_1$ deformation.
However, as we will see soon, setting $v^-=0$, one can take the limit $c_{1-}\to 0$ in the gauge-fixed Hamiltonian density, and this simplifies the flow equations drastically.

9. Transformation, $\cT_{\widetilde JJ}$, generating the $\widetilde JJ$ deformation 
\bal\la{tjj}
\cT_{\widetilde JJ}:\quad &x^1\to {1\ov A_{11}}x^1\,,\quad p_1\to A_{11}p_1\,,\quad \tilde X_1\to A_{11}\tilde X_1\,,\quad V^1\to {1\ov A_{11}} V^1\,,\\
&\Upsilon^1\to {1\ov A_{11}}\Upsilon^1\,,\quad \cU_1\to A_{11}\cU_1\,,
\eal
\bal\nn
A_{11}{\pa\cH_{A_{11}}\ov\pa A_{11}} = {\widetilde JJ}\,,\quad \widetilde JJ\equiv \widetilde J^0J^1 -\widetilde J^1J^0 = -\e_{\a\b}\widetilde J^\a J^\b\,.
\eal
The rescaling of  $\tilde X_1$, $V^1$ and $\Upsilon^1$, $\cU^1$ is necessary for a gauge-fixed Hamiltonian density to depend on the differences $x'^1-v^1$ and $p_1-u_1$.

10. Transformation, $\cT_{ T^1{}_1}$, generating the $T^1{}_1$ deformation 
\bal\nn
\cT_{ T^1{}_1}:\quad x^-\to {1\ov A_{--}}x^-\,,\quad p_-\to p_- A_{--}\,,
\eal
\bal\nn
 A_{--}{\pa\cH_{A_{--}}\ov\pa A_{--}} = T^1{}_1\,.
\eal
Due to the light-cone gauge condition $p_-=1$ the deformation with $A_{--}$ is equivalent to a change of the circumference of the cylinder the gauge-fixed model lives on. 

\medskip

Now, we can generate a ten-parameter transformation by using the following sequence of the one-parameter ones
\bal\la{transfA}
\cT_A=\cT_{ T^1{}_1}\cT_{\widetilde JJ}\cT_{\widetilde JT_1}\cT_{ \widetilde J^1}\cT_{ JT_0}\cT_{ J^1}\cT_{ T^1{}_0}\cT_{ JT_1}\cT_{\widetilde JT_0}\cT_{T\overline T}\,.
\eal
We see that to generate all these deformations we  need all three pairs of the auxiliary fields 
$(V^a_\a\,,\, \tilde X_a)$ and the pair $(\cU_{1\a}\,,\, \Upsilon^1)$. Thus, in what follows we set 
$$\cU_{+\a} =0\,,\, \Upsilon^+=0\,,\quad\cU_{-\a}=0\,,\, \Upsilon^-=0\,.$$

\medskip

The transformation \eqref{transfA} can be represented by a single canonical transformation. 
Introducing a column  of coordinates
\bal\nn
\cX = \big\{\cX^{\hat 1}\, \cX^{\hat 2},\ldots, \cX^{\hat 7}  \big\}=\big\{x^+,x^-,x^1,\tilde X_+,\tilde X_-,\tilde X_1,\Upsilon^1\big\}\,,
\eal
and a column of momenta
\bal\nn
\cP= \big\{\cP_{\hat 1},\cP_{\hat 2},\ldots, \cP_{\hat 7}  \big\}=\big\{p_+,p_-,p_1,V^+,V^-,V^1,\cU_1\big\}\,,
\eal
the transformation can be written in the form\footnote{We use $\cX'$ to write the canonical transformation because it is nonlocal in terms of $\cX$. 
The nonlocality does not cause any problem because the constraints in \eqref{S2} depend only on $\cX'$. }
\bal\la{cantransf}
\cX'\to A_x\cX' + B_x \cP +C_x\,,\quad \cP\to A_p^T\cP + B_p^T\cX' +C_p\,.
\eal
Here the matrices $A_x$, $B_x$ , $A_p $ and $B_p$, and the columns $C_x$, $C_p$ satisfy the identities
\bal\nn
A_xA_p+B_xB_p=\bI\,,\quad A_xB^T_x+B_xA_x^T=0\,,\quad A_p^TB_p+B_p^TA_p=0\,,
\eal
which are necessary for the transformation to be canonical, and additional identities
\bal\nn
A_pA_x+B_x^TB_p^T&=\bI\,,\quad A_pB_x+(A_pB_x)^T=0\,,\quad B_pA_x+(B_pA_x)^T=0\,,\\
C_p^TA_x\cX'+\cX'^TB_pC_x&=-{c_{1-}\ov A_{11}}x'^1\,,\quad C_pB_x\cP + \cP^TA_pC_x=0\,,\quad C_p^TC_x=0\,,
\eal
which lead to $\cP^T \cX' \to \cP^T \cX'  -{c_{1-}\ov A_{11}}x'^1$.
The explicit form of the matrices can be found in appendix \ref{canmatr}.

%%%%%%%%%%%%%%%%%%%%%%%%%%%%%%%%%%%%
\section{Deformed Hamiltonian and the flow equations}

In this section we use the ten-parameter canonical transformation \eqref{cantransf}, and the light-cone gauge conditions \eqref{ulc} imposed on the transformed coordinates and momenta to derive the gauge-fixed Hamiltonian density $\cH_A$ which we interpret as a deformation of the Hamiltonian density $\cH_0$. We then derive the ten-parameter flow equations for the Hamiltonian density $\cH_A$ which generalise the one-parameter equations discussed in section \ref{cantransform}. Then, these equations are converted into the flow equations for the spectrum.
Finally, we find level-matching conditions physical states must obey, and derive the
homogeneous inviscid Burgers equations for the energy of those states.

%%%%%%%%%%%%%%%%%%%%%%%%%%%%%%%%%%%%
\subsection{Deformed Hamiltonian}

Now, we can derive the gauge-fixed Hamiltonian density $\cH_A$ which we consider as a deformation of $\cH_0$. 
First, we note that up to total derivatives  the kinetic term in the action (\ref{S2}) does not change under the canonical transformation \eqref{cantransf}. 
Then, as was mentioned before on the constraints surface  the constraint $C_1$  is equivalent to  $\tilde C_1$ \eqref{tC1}. Clearly,
under  \eqref{cantransf}, $\tilde C_1$ changes as  
\bal\nn
\tilde C_1\overset{\cT_A}{\longrightarrow} \tilde C_1= \cP^T \cX' -{c_{1-}\ov A_{11}}x'^1+p_kx'^k\,.
\eal
Then, we obviously have
\bal\nn
C_a^V\overset{\cT_A}{\longrightarrow}  C_a^V= (A_x\cX' + B_x \cP+C_x)^{\hat 3+\hat a} = (A_x)^{\hat 3+\hat a}{}_{r}\cX'^r + (B_x)^{\hat 3+\hat a, r} \cP_r +C_x^{\hat 3+\hat a}\,.
\eal
\bal\nn
C^1_\cU\overset{\cT_A}{\longrightarrow}  C^1_\cU= (A_x\cX' + B_x \cP)^{\hat 7} = (A_x)^{\hat 7}{}_{r}\cX'^r + (B_x)^{\hat 7,r} \cP_r\,,\quad r=\hat1, \hat2,\ldots,\hat7\,.
\eal
Finally, a simple calculation gives the transformed Virasoro constraint $C_2$
\bal\la{C2f}
C_2\overset{\cT_A}{\longrightarrow}  C_2&={1\ov2}\tilde G^{rs} \cP_r \cP_s + {1\ov2}\cX'^r \cX'^s \tilde G_{rs}+\tilde G^{rk} \cP_r p_k + \cX'^r x'^k \tilde G_{rk}\\
&+\hat G^{r}{}_{s} \cP_r \cX'^s +\check G^{r}{}_{k} \cP_r x'^k + \cX'^r p_k \check G_{r}{}^{k}+\tilde\cH_x\,.
\eal
The coefficients in this equation are listed  in appendix \ref{virc}.

\medskip

Now, we  impose the light-cone gauge conditions \eqref{ulc} on the transformed coordinates and momenta. Then, from the analysis of 
the canonical transformation \eqref{cantransf} we know that the zero mode $v^-$ does not generate any of the deformations. Thus, in addition to  \eqref{ulc} we can set it to 0 (and we have already set $\cU_{\pm\a} =\Upsilon^\pm=0$)
\bal\nn
v^-=0\,.
\eal
Next, we  solve the constraints $\tilde C_1=0$, $C_a^V =0$ and $C^1_\cU=0$ for 
$$x'^-\,,\, \tilde X'_+\,,\, \tilde X'_-\,,\, \tilde X'_1\,,\, \Upsilon'^1\,.$$
The solution can be written in the  form
\bal\la{cxpr}
\cX'^r=\vk^r \, p_+ + \chi^r \,,
\eal
where we also have $\cX'^{\hat1}=x'^+=0$ and $\cX'^{\hat3}=x'^1$. The coefficients $\vk^r$ and $\chi^r$ can be read off from eqs.(\ref{cxp2}).

\medskip

Using the solution, 
we bring $C_2$ to the form $
C_2 = \cG_{2}\, p_+^2  +  \cG_{1}\, p_+  + \cG_{0}
$,
where the coefficients $\cG_{i}$  are listed  in appendix \ref{virc}.
One then checks that even though the  transformation  \eqref{cantransf} is singular in the limit $c_{1-}\to 0$,  the coefficients $\cG_{i}$ are regular at $c_{1-}=0$. In fact it is necessary to have $v^-=0$ for the regularity.  Thus, in what follows we set $c_{1-}=0$, and only discuss this case. If $c_{1-}\neq 0$ then there should exist a redefinition of the parameters which makes the gauge-fixed Hamiltonian independent of $c_{1-}$. 

\medskip

The solution of  the quadratic equation 
\bal \la{eqc2}
C_2 &= \cG_{2}\, p_+^2  +  \cG_{1}\, p_+  + \cG_{0} =0\, \,,
\eal 
which reduces to $-\cH_0$ when all the parameters vanish gives us  the deformed Hamiltonian  and action
\bal \la{S4} 
S_A=\int_{0}^{R}\, {\rm d}\s{\rm d}\tau\, \left( p_\mu \dot{x}^\mu
\,-\, \cH_A \right)\, ,\quad \cH_A \,=\, -p_+(p_\mu, x^\mu ,x'^\mu ) \,,\quad R={P_-}\,.
 \eal
We consider some deformations of a model with the Hamiltonian  
\eqref{ch0} in appendix \ref{examples}.

%\medskip

%%%%%%%%%%%%%%%%%%%%%%%%%%%%%%%%%%%%
\subsection{Flow equations for the density of the Hamiltonian}

To simplify the understanding of the origin of various terms in the flow equations in this subsection we  use the notations 
\bal\la{aTTb}
\aTTb \equiv a_{+-}\,,\quad \aJtTt \equiv -a_{+1}\,,\quad \aJTs \equiv a_{1-}\,,\quad \aJTt \equiv b_{+1}\,,\quad \aJtTs \equiv b_{--}\,,\quad v\equiv -v^+\,,
\eal
so that the flow equations at the leading order would have the form
\bal\nn
{\pa\cH_A\ov \pa \a_\cO} = \cO\,,\quad {\pa\cH_A\ov \pa v} = T^1{}_0\,.
\eal
 
 The Hamiltonian density $\cH_A$ (or, better to say the defining equation \eqref{eqc2}, see appendix \ref{flow-eq}) can be used to derive the flow equations with respect to the ten deformation parameters.
First of all, it is easy to check that the coefficients $\cG_{i}$, and, therefore, $\cH_A$, depend on the differences $x'^1-v^1$ and $p_1-u_1$. Thus, 
\bal\la{eqvu}
{\pa\cH_{A}\ov\pa v^1} = J^1\,,\quad 
{\pa\cH_{A}\ov\pa u_1} = \widetilde J^1\,.
\eal
Moreover, all the operators appearing on the right hand side of the flow equations also must depend on  $x'^1-v^1$ and $p_1-u_1$. Let us introduce the following improved  currents $\bJ^\a$, $\widetilde \bJ^\a$ which depend on those differences  
\bal\la{imprJ0}
\bJ^0=p_1-u_1 = J^0-u_1\,,\quad \bJ^1=J^1\,,
\\
\widetilde\bJ^0=x'^1-v^1=\widetilde J^0-v^1\,,\quad \widetilde\bJ^1=\widetilde J^1\,,
\eal 
and  the improved stress-energy tensor $\bT^\a{}_\b$
\bal\nn
\bT^0{}_0=T^0{}_0 \,,\quad \bT^1{}_0=T^1{}_0\,,
\eal 
\bal\la{imprT01}
\bT^0{}_1&= (p_1-u_1) (x'^1-v^1)+p_kx'^k -v\, T^0{}_0 =T^0{}_1-v\, T^0{}_0 -v^1J^0-u_1\widetilde J^0+u_1v^1\,,
\\
\bT^1{}_1&=\cH_A-{\pa \cH_A\ov \pa x'^k} x'^k   -p_k {\pa \cH_A\ov \pa p_k}-{\pa \cH_A\ov \pa x'^1} (x'^1-v^1) -(p_1-u_1) {\pa \cH_A\ov \pa p_1}-v\, T^1{}_0\\
&= T^1{}_1 -v\, T^1{}_0-v^1J^1-u_1\widetilde J^1\,,
\eal 
whose components $\bT^\a{}_1$ are also shifted by  $v\, T^\a{}_0$.

Then, in terms of the improved currents (\ref{imprJ0}, \ref{imprT01}) the flow equations take the form (in the limit $c_{1-}\to 0$)
\bal\la{eqvp}
{\pa\cH_{A}\ov\pa v} = \bT^1{}_0= T^1{}_0\,,
\eal
\bal\la{eqamm}
A_{--}{\pa\cH_{A}\ov\pa A_{--}} = \bT^1{}_1= T^1{}_1-vT^1{}_0-v^1J^1-u_1\widetilde J^1\,,
\eal
\bal\la{eqa11}
A_{11}{\pa\cH_{A}\ov\pa A_{11}} &=\widetilde\bJ\bJ=\widetilde JJ -u_1 J^1+v^1\widetilde J^1\,,
\eal
\bal\la{eqapm}
A_{--}{\pa\cH_{A}\ov\pa \aTTb} = \bT\overline\bT\,,
\eal
\bal\la{eqa1m}
{A_{--}}{\pa\cH_{A}\ov\pa \aJTs} &= A_{11}\bJ\bT_1\,,
\\
{\pa\cH_{A}\ov\pa \aJTt} &= A_{11}\bJ\bT_0+{\aJtTs\ov A_{--}}\bT\overline\bT
\,,
\\
A_{--}{\pa\cH_{A}\ov\pa \aJtTs} &= {1\ov A_{11}}\widetilde\bJ\bT_1\,,
\\
{\pa\cH_{A}\ov\pa \aJtTt} &= {1\ov A_{11}}\widetilde\bJ\bT_0+{\aJTs\ov A_{--}}\bT\overline\bT\,.
\eal

A set of nine-parameter flow equations similar to eqs.(\ref{eqvp}-\ref{eqa1m}) with $A_{--}=1$ was recently proposed in \cite{Mezei2019a}, see eq.(4.1) there. The equations in \cite{Mezei2019a} were conjectured to be universal and valid for any model. It is immediately seen, however, that they
disagree with  our flow equations. An obvious difference is that the coefficients in front of the 
$\bT\overline\bT$ operator on the r.h.s. of our flow equations \eqref{eqa1m} depend on the deformation parameters $\aJtTs$ and $\aJTs$ of the quadratic operators $\widetilde\bJ\bT_1$ and $\bJ\bT_1$ while the r.h.s. of the flow equations of \cite{Mezei2019a} do not depend on the parameters of quadratic operators at all. A more careful look at the flow equations of  \cite{Mezei2019a}, see Table 2 there, shows that in their equations $\bT\overline\bT$ only appears  in the equation for the \TTb\ deformation. Moreover, if one switches off their parameters $a$, $\bar a$, $b$ of linear deformations then their equations coincide with the flow equations for one-parameter deformations.  The appearance of the $\bT\overline\bT$ operator on the r.h.s. of \eqref{eqa1m}, however, is natural, and reflects the noncommutativity of some of the one-parameter flows generated by the quadratic operators.

The reason for the disagreement appears to be quite interesting. The authors of \cite{Mezei2019a} only analysed the model of a single massless free scalar which is a CFT with left- and right-moving conserved currents.  As we show in section 4, for such a model there is a choice of the \TTb\ deformation parameter $\aTTb$ such that the deformed Hamiltonian does not depend on the parameters of the $\rJ \T$ and $\bar\rJ \overline{\T}$ deformations. 
The flow equations with respect to these parameters then become 
extra constraints \eqref{eqbep3c}  which generalise the relations $\rJ \T=0$ and $\bar\rJ \overline{\T}=0$ of the undeformed CFT. These constraints were missed in \cite{Mezei2019a}.
Their existence breaks the uniqueness of the flow equations because these constraints can be added to the r.h.s. of the flow equations with arbitrary coefficients. In particular,
 one can make a choice of the deformation parameters which removes the 
$\bT\overline\bT$ terms from the flow equations \eqref{eqa1m}. Such a choice exists only for a CFT with left- and right-moving conserved currents, and for any other model one has to deal with the universal flow equations (\ref{eqvp}-\ref{eqa1m}).

\medskip

A simple analysis of eqs. \eqref{eqvu},  \eqref{eqvp} and  \eqref{eqamm} shows that the following rescaling of $v^1, v,u_1$ 
\bal\nn
v^1\to A_{--}v^1\,,\quad v\to A_{--}v\,,\quad u_1\to A_{--}u_1\,,\quad 
\eal
brings the equations to the  form
\bal\la{eqvu2}
A_{--}{\pa\cH_{A}\ov\pa v^1} = J^1\,,\quad 
A_{--}{\pa\cH_{A}\ov\pa u_1} = \widetilde J^1\,,\quad A_{--}{\pa\cH_{A}\ov\pa v} = T^1{}_0\,,\quad A_{--}{\pa\cH_{A}\ov\pa A_{--}} = T^1{}_1\,.
\eal
This rescaling, however, breaks the dependence of $\cH_A$ on  $x'^1-v^1$ and $p_1-u_1$. To restore the dependence we can rescale the world-sheet space coordinate $\s$
\bal\nn
\s\to\s/A_{--}\ \Rightarrow \ S_A\to S_A=\int_{0}^{RA_{--}}\, {{\rm d}\s\ov A_{--}}{\rm d}\tau\, \left( p_\mu \dot{x}^\mu
\,-\,\cH_A(p_\mu, x^\mu, A_{--}x'^\mu) \right)\,.
\eal
We see that to restore the canonical Poisson structure we also need to  rescale $p_\mu$
\bal\nn
p_\mu\to A_{--}p_\mu\ \Rightarrow \ S_A\to S_A=\int_{0}^{RA_{--}}\, {\rm d}\s{\rm d}\tau\, \left(p_\mu \dot{x}^\mu
\,-\, {1\ov A_{--}}\cH_A(A_{--} p_\mu, x^\mu ,A_{--}x'^\mu ) \right)\, .
\eal
One can then check by using the coefficients $\cG_{i}$ that 
\bal\nn
{1\ov A_{--}}\cH_A(A_{--} p_\mu, x^\mu ,A_{--}x'^\mu ) = \cH_A(p_\mu, x^\mu,x'^\mu )\big|_{A_{--}=1}\,.
\eal
Thus, without loss of generality one can set $A_{--}=1$, and consider $R$ as the tenth deformation parameter. In quantum theory the flow equation \eqref{eqamm} with respect to $A_{--}$  would imply the following flow equation for the energy with respect to $R$
\bal\nn
{\pa E_n\ov \pa R} = \bra n|T^1{}_1|n\ket\,,
\eal
where $|n\ket$ is an eigenstate of $H_A$ with the energy $E_n$.

\medskip

The $\widetilde JJ$ deformation can be also easily understood. It is clear from the way it was generated, 
see eq.\eqref{tjj}, and can be checked explicitly by using the defining equation \eqref{eqc2} that the deformed Hamiltonian  depends on $A_{11}$ only through the combinations
\bal\nn
A_{11}(p_1 - u_1)\,,\quad {1\ov A_{11}}(x'^1-v^1)\,.
\eal
Thus, the energy of a deformed model with parameters $A_{11}$, $u_1$, $v^1$ is related to the energy of the deformed model with  $A_{11}=1$ as 
\bal\nn
E_n(A_{11},P_1-Ru_1,\widetilde P^1-Rv^1) = E_n(1,A_{11}(P_1-Ru_1),(\widetilde P^1-Rv^1)/A_{11})\,,
\eal
where  $\widetilde P^1=x^1(R)-x^1(0)$ is basically an effective range of $x^1$ which may depend on winding numbers of several scalars and target space-time metric of the model, see the discussion below eq.\eqref{L0}. This can be also seen from the flow equations. Indeed, rescaling $u_1$ and $v^1$ as $u_1\to u_1/A_{11}$, $v^1\to A_{11} v^1$, and then rescaling the currents as $J^\a\to J^\a/A_{11}$, $\widetilde J^\a\to A_{11}\widetilde J^\a$ removes completely the $A_{11}$ dependence from the flow equations. Thus,  we may  set $A_{11}=1$.

%\medskip

%%%%%%%%%%%%%%%%%%%%%%%%%%%%%%%%%%%%
\subsection{Flow equations for the energy} \la{fleqen}

Since the Hamiltonian $\bfH_A$, the world-sheet momentum $\bfP$, the U(1) charge $\bfP_1$, and the dual charge $\widetilde \bfP^1$ are mutually commuting, 
the energy $E_n$ of their common eigenstate $|n\ket$ is a function of $R$, and the eigenvalues $P$, $P_1$, and $\widetilde P^1$ of $\bfP$, $\bfP_1$,  and $\widetilde \bfP^1$, and the deformation parameters.  
Then, the flow equations for the Hamiltonian density, and the Hellmann-Feynman
theorem lead to the following relations
\bal\la{nT11n1}
\bra n|T^0{}_0|n\ket &= {E_n\ov R}\,,\quad \bra n|T^0{}_1|n\ket =- {P\ov R}\,,\quad \bra n|J^0|n\ket = {P_1\ov R}\,, \quad \bra n|\widetilde J^0|n\ket = {\widetilde P^1\ov R}\,,
\\
\bra n|T^1{}_1|n\ket&={\pa E_n\ov \pa R} \,,\quad \bra n|T^1{}_0|n\ket={1\ov R}{\pa E_n\ov \pa v} \,,\quad \bra n|J^1|n\ket={1\ov R}{\pa E_n\ov \pa v^1} \,,\quad \bra n|\widetilde J^1|n\ket={1\ov R}{\pa E_n\ov \pa  u_1} \,.
\eal
In addition, assuming factorisation, for an operator of the form 
$\cO_{K_1K_2}=-\e_{\a\b}K^\a_1K^\b_2$, one has
\bal\nn
\bra n|\cO_{K_1K_2}|n\ket = \bra n|K^0_1|n\ket \bra n|K^1_2|n\ket- \bra n|K^1_1|n\ket \bra n|K^0_2|n\ket \,,
\eal
where $K^\a_i$ are conserved currents, and $|n\ket$ is an eigenvector of their charges.
By using these relations, it is straightforward to derive  flow equations for the energy of the state. 

\medskip

There is, however, a better form of  flow equations for the energy.
The Hamiltonian density   depends on  $p_1-u_1=J^0 -  u_1$, and $x'^1-v^1=\widetilde J^0 -v^1$, and therefore it is more natural to consider the energy $E_n$ as a function of the improved charges $\bP_1=P_1 - R u_1$, and $\widetilde\bP^1=\widetilde P^1 -Rv^1$. 
To make sure that the energy would depend on $u_1$ and $v^1$ only through $\bP_1$ and $\widetilde\bP^1$, one also should consider  $E_n$ as a function of 
\bal\nn
\check P \equiv P + {P_1\widetilde P^1\ov R}\,.
\eal
Indeed,  the world-sheet momentum can be represented as 
\bal\nn
P = -\int_0^Rd\s\, p_\mu x'^\mu  = -\int_0^Rd\s\, \big(({P_1\ov R}+\check p_1) ({\widetilde P^1\ov R}+\check x'^1)  +p_k x'^k\big)=- {P_1\widetilde P^1\ov R}+\check P\,,
\eal
where $\check p_1$ and $\check x'^1$ satisfy $\int_0^R \check p_1=0$, $\check x^1(R)=\check x^1(0)$. Clearly, this formula demonstrates the splitting of the world-sheet momentum into its intrinsic part $\check P$, and the part due to the rotation and winding in the $x^1$ direction.

\medskip

Thus, we can consider the energy as a function of $R$, $\check P$, $\bP_1$ and $\widetilde\bP^1$, and the deformation parameters $\a_\cO$, $v$, $A_{11}$
\bal\nn
E_n = E_n(R\,,\check P\,, \bP_1\,,\widetilde\bP^1\,,\a_\cO\,, v\,, A_{11})= E_n(R\,,\check P\,, P_1 - R u_1\,,\widetilde P^1 -Rv^1\,,\a_\cO\,, v\,, A_{11})\,.
\eal
It is now clear that the last two relations in \eqref{nT11n1} take the form
\bal\nn
\bra n|J^1|n\ket=-{\pa E_n\ov \pa \widetilde \bP^1} \,,\quad \bra n|\widetilde J^1|n\ket=-{\pa E_n\ov \pa  \bP_1} \,.
\eal
Then, the  flow equations for the Hamiltonian density depend on the improved currents and stress-energy tensor  (\ref{imprJ0}, \ref{imprT01}), and their expectation values are
\bal\la{nT00n2}
 \bra n|\bT^0{}_1|n\ket &=- {1\ov R}\big(\check P-{\bP_1\widetilde \bP^1\ov R}+ vE_n\big)\,,\quad \bra n|\bT^1{}_1|n\ket={\pa E_n\ov \pa R} -{v\ov R}{\pa E_n\ov \pa v} \,,
\\
\bra n|\bJ^0|n\ket &= {\bP_1\ov R}\,, \quad \bra n|\widetilde \bJ^0|n\ket = {\widetilde \bP^1\ov R}\,.
\eal

\medskip

By using these relations, one gets the following flow equations for the energy of the state $|n\ket$ (with $A_{--}=1$)
\bal\la{eqa11en}
A_{11}{\pa E_n\ov\pa A_{11}} &= \bP_1 {\pa E_n\ov\pa \bP_1} -\widetilde\bP^1 {\pa E_n\ov\pa \widetilde\bP^1}  \,,
\eal
\bal\la{eqapmen}
{\pa E_n\ov\pa \aTTb} &= -E_n{\pa E_n\ov \pa R} - {\check P-{\bP_1\widetilde \bP^1\ov R}\ov R}{\pa E_n\ov \pa v}\,,
\\
 {1\ov A_{11}}{\pa E_n\ov\pa \aJTs} &=\bP_1\big({\pa E_n\ov \pa R}- {v\ov R}{\pa E_n\ov \pa v}\big)-\big(\check P-{\bP_1\widetilde \bP^1\ov R}+v E_n\big){\pa E_n\ov \pa\widetilde \bP^1}\,,
\\
 {1\ov A_{11}}{\pa E_n\ov\pa \aJTt} &={\bP_1\ov R}{\pa E_n\ov \pa v}+ E_n{\pa E_n\ov \pa\widetilde \bP^1}+{\aJtTs\ov A_{11}}{\pa E_n\ov\pa \aTTb}
\,,
\\
A_{11}{\pa E_n\ov\pa \aJtTs} &=\widetilde\bP^1\big({\pa E_n\ov \pa R}- {v\ov R}{\pa E_n\ov \pa v}\big)-\big(\check P-{\bP_1\widetilde \bP^1\ov R}+v E_n\big){\pa E_n\ov \pa \bP_1}\,,
\\
A_{11}{\pa E_n\ov\pa \aJtTt} &= {\widetilde\bP^1\ov R}{\pa E_n\ov \pa v}+ E_n{\pa E_n\ov \pa \bP_1}+A_{11}\aJTs{\pa E_n\ov\pa \aTTb}\,.
\eal
It is unclear how to solve this system of equations. However, the equation \eqref{eqa11en} with respect to $A_{11}$ for the $\widetilde JJ$ deformation can be easily solved. Its solution is
\bal\la{eqa11ensol}
E_n(\bP_1\,,\widetilde\bP^1\,,A_{11}) = E_n(A_{11}\bP_1\,,{\widetilde\bP^1\ov A_{11}}\,,1)\,,
\eal
where we kept the energy dependence only on the essential variables. Thus, without loss of generality we can set $A_{11}=1$ in the remaining five equations.

%%%%%%%%%%%%%%%%%%%%%%%%%%%%%%%
\subsection{Homogeneous inviscid Burgers equation}\la{spectrum}

The deformed and undeformed models are obtained from the same extended string sigma model by applying different light-cone type gauge conditions. 
 The physical quantities, therefore, must be the same in both models.  However, as usual in a light-cone gauge, physical states must satisfy additional conditions which follow from the requirement that the strings are closed. We will refer to these conditions as to generalised level-matching conditions. In this section we determine which  states of the deformed model are physical, and discuss a relation between  the energy of physical states in the deformed and undeformed models. We set $A_{--}=1$ and use $R$ as a  parameter.

\medskip

Let us denote $x^a_0$, $\tilde X_a^0$, $\Upsilon^a_0$, $p_a^0$, $V^a_0$ and $\cU_a^0$ the coordinates and momenta of the extended string model \eqref{S2} before one applies the canonical transformation \eqref{cantransf}. They are related to the transformed coordinates and momenta as 
in eqs.\eqref{cantransf}. The transformed coordinates and momenta satisfy the light-cone gauge conditions and  the constraints. Substituting the solution of the  gauge conditions and  the constraints into the relations \eqref{cantransf}, one finds for the coordinates
\bal\la{xpsol}
 x'^-_0&=-\bT^0{}_1\,,\\
 x'^1_0&= \aJTs \bT^0{}_1+{ \widetilde\bJ^0 \ov A_{11}} + \aJTt \cH_A\,,\\
 x'^+_0&=(\aTTb +\aJTs \aJtTt +  \aJtTs \aJTt) \bT^0{}_1  + 
 \aJtTt \big({\widetilde\bJ^0\ov A_{11}} + \aJTt\cH_A\big)  + A_{11} \aJTt \bJ^0 + v\,,\\
 \tilde X'_a{}^0&=0\,,\quad \Upsilon'{}^a_0 =0\,,\quad a= +,-,1\,,
\eal
and for the momenta
\bal\la{psol}
p_+^0&=-\cH_A\,,\\
 p_-^0&=1 -\aTTb \cH_A+ A_{11} \aJTs \bJ^0 + 
 \aJtTs { \widetilde\bJ^0 \ov A_{11}}+ \aJTs \aJtTs \bT^0{}_1  \,,\\
 p_1^0&= A_{11} \bJ^0 + \aJtTt\cH_A+\aJtTs \bT^0{}_1 \,,\\
 V^+_0&=-v\,,\\
 V^1_0&=v^{1}/A_{11} - \aJTt \cH_A\,,\\
\cU_1^0&=v^{1}/A_{11} - \aJTt A_{11} u_{1}\cH_A\,,
\eal
where the improved currents and stress-energy tensor are given by  (\ref{imprJ0}, \ref{imprT01}).

Eqs.\eqref{xpsol} show that the generalised level-matching conditions  are
\bal\la{physcond}
x^-_0(R)-x^-_0(0)&=-\int_0^R d\s\, \bT^0{}_1 = \bP =0\,,\\
x^1_0(R)-x^1_0(0)&=\widetilde P_0^1=\int_0^R d\s\, ({ \widetilde\bJ^0 \ov A_{11}} + \aJTt \cH_A) = { \widetilde\bP^1 \ov A_{11}} + \aJTt E_A = w \cR^1_0\,,\\
x^+_0(R)-x^+_0(0)&=\aJtTt w \cR^1_0+\int_0^R d\s\, (A_{11} \aJTt \bJ^0 + v) = \aJtTt w \cR^1_0+ A_{11} \aJTt \bP_1 + vR = 0\,,
\eal
where $w\in\bZ$ is the winding number of $x^1_0$ and $x^1$, and the improved world-sheet momentum $\bP$, and charges $\bP_1$ and $\widetilde\bP^1$ are given by
\bal
\bP&=P+v\, E +v^1P_1+u_1\widetilde P^1-u_1v^1R\,,\\
\bP_1&=P_1-u_1R\,,\\
\widetilde\bP^1 &= \widetilde P^1 -v^1R = w \cR^1 - v^1R\,.
\eal
Solving the generalised level-matching conditions  for $P$, $v$ and $v^1$, one gets
 \bal\la{physcond2}
 P &= w\big( {\cR_0^1\ov R} \aJtTt\, E_A +{A_{11}\cR^1_0\ov R}\bP_1  - {\cR^1\ov R}P_1 \big)\,,\\
 vR&= -\aJtTt w \cR^1_0- A_{11} \aJTt \bP_1  \,,\\
 v^1R&=A_{11}\aJTt E_A +w(\cR^1-A_{11}\cR^1_0)\,.
 \eal
 
 Having found the  generalised level-matching conditions, we can now relate the energies of physical states of the deformed and undeformed models. 
 Integrating the relations \eqref{psol} over $\s$, one gets
 \bal\la{Erel}
E_0&=E_A \,,\\
 R_0&=R -\aTTb E_A + 
 \aJtTs { \widetilde\bP^1 \ov A_{11}}+ \aJTs(A_{11} \bP_1 - \aJtTs \bP ) \,,\\
 P_1^0&= A_{11} \bP_1-\aJtTs \bP + \aJtTt E_A  \,,
\eal
where $R_0\equiv P_-^0 = \int_0^R p_-^0$ is the circumference of the cylinder, $P_1^0= \int_0^R p_1^0$ is the $U(1)$ charge, and  $E_0$ is the energy 
  of the undeformed model which is equal to the energy of the deformed model for physical states.  
 These equations generalise the integrated form of the homogeneous inviscid Burgers equation which describes the spectrum of a \TTb\ deformed model for states with vanishing world-sheet momentum. For physical states one can set $\bP=0$ and ${ \widetilde\bP^1}  = w \cR^1-A_{11}\aJTt E_A$ but we prefer to keep them in \eqref{Erel} because they might be important for states which do not satisfy the  generalised level-matching conditions. Due to the momentum $P_1$ quantisation 
 \bal\nn
 P_1^0={2\pi m\ov \cR^1_0}\,,\quad P_1={2\pi m\ov \cR^1}\,,\quad m\in\bZ\,,
 \eal
 and the last equation in \eqref{Erel} can be considered as a relation between the ranges $\cR_0^1$ and $\cR^1$ of $x^1_0$ and $x^1$.

%%%%%%%%%%%%%%%%%%%%%%%%%%%%%%%%%%%%
\section{Flow equations in CFT conventions}

To analyse the deformations of a CFT it is convenient to use the light-cone world-sheet coordinates 
\bal\nn
\s^\pm = \s\pm\tau\,,\quad \pa_\pm = {1\ov2}(\pa_\s\pm\pa_\tau)\,,\quad \pa_\a J^\a= \pa_+ J^++\pa_-J^-\,.
\eal
 The light-cone components of a vector $V^\a$, a co-vector $U_\a$, and the stress-energy tensor $T^\a{}_\b$  are related to the $0,1$ components as
 \bal\nn
 V^\pm &= V^1\pm V^0\,,\quad V^0={1\ov2}(V^+-V^-)\,,\quad V^1={1\ov2}(V^++V^-)\,,
 \\
 U_\pm &= {1\ov2}(U_1\pm U_0)\,,\quad U_0=U_+-U_-\,,\quad U_1=U_++U_-\,,
 \\
 T^+{}_+&= {1\ov2}(T^0{}_0+T^1{}_1+T^0{}_1+T^1{}_0)\,,\quad T^-{}_-= {1\ov2}(T^0{}_0+T^1{}_1-T^0{}_1-T^1{}_0)\,,
 \\
 T^-{}_+&= {1\ov2}(T^1{}_1-T^0{}_0-T^0{}_1+T^1{}_0)\,,\quad T^+{}_-= {1\ov2}(T^1{}_1-T^0{}_0+T^0{}_1-T^1{}_0)\,.
 \eal
 In a CFT one has $T^0{}_0+T^1{}_1=0$, $T^0{}_1+T^1{}_0=0$, and therefore $T^+{}_+=T^-{}_-=0$. 
 In terms of the light-cone components the deformation operators become
 \bal\nn
 T\overline{T}&= -\e_{\a\b}T^\a{}_1T^\b{}_0= -2\e_{\a\b}T^\a{}_-T^\b{}_+=T^+{}_-T^-{}_+-T^-{}_-T^+{}_+\,,
 \\
 VT_1&= -\e_{\a\b}V^\a T^\b{}_1={1\ov2} \big(V^+ ( T^-{}_++T^-{}_- ) - V^- (T^+{}_++T^+{}_- )\big)\,,
 \\
 VT_0&= -\e_{\a\b}V^\a T^\b{}_0={1\ov2} \big(V^+ ( T^-{}_+-T^-{}_- ) - V^- (T^+{}_+-T^+{}_- )\big)\,,
 \\
 VT_\pm&= -\e_{\a\b}V^\a T^\b{}_\pm={1\ov2} \big(V^+ T^-{}_\pm - V^- T^+{}_\pm\big)\,,
 \eal
 where $V^\a$ could be either $J^\a$ or $\widetilde J^\a$.
 
 %%%%%%%%%%%%%%%%%%%%%%%%%%%%%%%%%%%%
\subsection{Flow equations for the Hamiltonian density}\la{fleqHdcft}
 
 Introducing new deformation parameters (and setting $A_{--}=1$)
 \bal\nn
 \a_{\pm}=\aJTs\pm \aJTt\,,\quad \tilde\a_{\pm}=\aJtTs\pm \aJtTt\,,
 \eal
 we rewrite the flow equations (\ref{eqa1m}) in the form
\bal\la{eqalpm}
{\pa\cH_{A}\ov\pa \a_{\pm}} &=A_{11}\bJ\bT_\pm\pm\, {\tilde\a_{+}+\tilde\a_{-}\ov4}\bT\overline\bT\,,
\\
{\pa\cH_{A}\ov\pa \tilde\a_{\pm}} &= {1\ov A_{11}}\widetilde\bJ\bT_\pm
\pm {\a_{+}+\a_{-}\ov4}\bT\overline\bT\,.
\eal
For a generic model (even a CFT) there is no reason to mix the currents $J$ and $\widetilde J$. 
However, the situation is different if we want to deform a Lorentz-invariant model with the Lagrangian
\bal\la{L0}
 \cL_0 = -{1\ov2}\pa_\a x^\mu \pa^\a x^\nu\,G_{\mu\nu} -V(x)\,,
 \eal
 which is invariant under shifts of $x^1$, and whose target-space metric components $G_{1\nu}$ are independent of $x$. Then, shifting and rescaling $x^1$
 \bal\nn
 x^1\to {1\ov \sqrt{G_{11}}}(x^1 -G_{1k}x^k)\,,
 \eal 
 we (almost) decouple $x^1$ from the other fields
 \bal\la{L00}
 \cL_0 \to \cL_0= -{1\ov2}\pa_\a x^1 \pa^\a x^1-{1\ov2}\pa_\a x^k \pa^\a x^l\,\tilde G_{kl} -V(x)\,.
 \eal
 We see that $x^1$ is a free massless scalar whose only knowledge of the other fields is in its winding numbers. For such a scalar the topological current $\widetilde J$ is dual to $J^\a = -\pa^\a x^1$
 \bal\nn
 \widetilde J^0 = x'^1=-J^1\,,\quad \widetilde J^1 = -\dot x^1=p_1=-J^0\,,
 \eal
and introducing
\bal\nn
\rJ_{\pm}^\a=-{1\ov2}( J^\a\pm  \widetilde J^\a)\,,\quad \rJ_{\pm}^0=\mp\pa_\pm x^1\,,\quad \rJ_{\pm}^1=\pa_\pm x^1\,,
\eal 
\bal\nn
\rJ_{+}^-=2\pa_+ x^1\,,\quad \rJ_{+}^+=0\,,\quad \rJ_{-}^+=2\pa_- x^1\,,\quad \rJ_{-}^-=0\,,
\eal 
we see that the current $\rJ^\a\equiv \rJ_{+}^\a$ is purely left-moving (holomorphic) while $\bar\rJ^\a\equiv\rJ_{-}^\a$ is purely right-moving (anti-holomorphic). Clearly, a generic deformation would ruin these nice properties but one can still hope that these currents $\rJ^\a$ and $\bar\rJ^\a$  would suit better to describe the deformation. In what follows discussing the deformation of a model with Lagrangian \eqref{L0}, we first decouple the scalar $x^1$, and then deform the model.

\medskip

Coming back to the flow equations (\ref{eqalpm}), we see that in the presence of the $\tilde JJ$-deformation the currents $\rJ^\a$ and 
$\bar\rJ^\a$ should be defined as
\bal\la{jpm}
\rJ_{\pm}^\a=-{1\ov2}(  {A_{11}}J^\a \pm {1\ov A_{11}}\widetilde J^\a)\,,\quad \tilde JJ = -2\e_{\a\b}\rJ_{+}^\a\rJ_{-}^\b\,.
\eal 
The appearance of $A_{11}$ in this formula is expected because the $\widetilde JJ$-deformation \eqref{tjj} rescales the coordinate $x^1$ and momentum $p_1$, and the factors of  $A_{11}$ just reverse this rescaling. 

\medskip

Now,  introducing the deformation parameters
\bal\nn
\b_\pm = -{1\ov2}(\a_\pm+\tilde\a_\pm)\,,\quad \bar\b_\pm =  -{1\ov2}(\a_\pm-\tilde\a_\pm)\,,
\eal
we bring the flow equations (\ref{eqalpm}) to the form
\bal\la{eqbepm}
{\pa\cH_{A}\ov\pa \b_{\pm}} =2\cJ\bT_\pm \pm\, {\b_{+}+\b_{-}\ov2}\bT\overline\bT\,,
\\
{\pa\cH_{A}\ov\pa \bar\b_{\pm}} = 2\bar\cJ\bT_\pm
\mp {\bar\b_{+}+\bar\b_{-}\ov2}\bT\overline\bT\,.
\eal
Here $\cJ^\a \equiv  \cJ^\a_+$ and $\bar\cJ\equiv  \cJ^\a_-$ are the improved currents
\bal\la{imjpm}
\cJ_{\pm}^\a=-{1\ov2}(  {A_{11}}\bJ^\a\pm {1\ov A_{11}} \tilde\bJ^\a)\,,
\eal 
and up to a normalisation in the Euclidean CFT terminology $J T_+ \leftrightarrow J\T$, $J T_- \leftrightarrow J\overline T$, 
$\bar J T_+ \leftrightarrow  \bar J T$, $\bar J T_- \leftrightarrow \bar J\overline \T$. 

\medskip

A nice feature of these equations is that in the absence of linear deformations ($v^1=v=u_1=0$) the flow equations for the left- and right-moving currents are coupled to each other only through the \TTb\ operator. Nevertheless, it  still may not be the best form of the equations because they all involve the \TTb\  operator. Indeed,  we have the flow equation for the \TTb\ deformation, and we can make nonlinear changes of the parameter $a_{+-}$. Performing
 the following  transformation 
\bal\nn
a_{+-}= \a+{\vk_{+-}\ov2} \b_{+}\b_{-}+{\vk_{+} \ov4}\b_{+}^2+{\vk_{-}\ov4} \b_{-}^2+{\bar\vk_{+-}\ov2} \bar\b_{+}\bar\b_{-}+{\bar\vk_{+}\ov4} \bar\b_{+}^2+{\bar\vk_{-}\ov4} \bar\b_{-}^2\,,
\eal
one finds that   the equations (\ref{eqapm}, \ref{eqbepm}) take the form
\bal\la{eqapm2}
{\pa\cH_{A}\ov\pa \a} = \bT\overline\bT\,,
\eal
\bal\la{eqbep2}
{\pa\cH_{A}\ov\pa \b_{+}} =2\cJ\bT_+\, +\, {\b_{+}(1+\vk_+)+\b_{-}(1+ \vk_{+-}) \ov2}\bT\overline\bT\,,
\\
{\pa\cH_{A}\ov\pa \b_{-}} =2\cJ\bT_-\, -\, {\b_{+}(1-\vk_{+-})+\b_{-}(1- \vk_{-}) \ov2}\bT\overline\bT\,,
\\
{\pa\cH_{A}\ov\pa \bar\b_{+}} = 2\bar\cJ\bT_+
- {\bar\b_{+}(1- \bar\vk_{+})+\bar\b_{-}(1-\bar\vk_{+-})\ov2}\bT\overline\bT\,,
\\
{\pa\cH_{A}\ov\pa \bar\b_{-}} = 2\bar\cJ\bT_-
+ {\bar\b_{+}(1+\bar\vk_{+-})+\bar\b_{-}(1+\bar\vk_{-})\ov2}\bT\overline\bT\,.
\eal
It is clear that natural values of the parameters $\vk$'s are $0,\pm1$. A particular choice of  $\vk$'s depends on the model under study. If the model does not have left- and right-moving conserved currents then the original choice with all $\vk$'s equal to 0 seems to be the best one. 
However, if a CFT has left- and right-moving  currents then there are better choices. 
Since in such a CFT $J T_+=0$ and $\bar J T_-=0$,
it seems reasonable to choose $\vk$'s so that the flow equations for $J T_-$ and $\bar J T_+$ would take the simplest form. 
Thus, one would choose  $\vk_-=1$, $\vk_{+-}=1$, $\bar\vk_+=1$, $\bar\vk_{+-}=1$. Then, the remaining two flow equations take the simplest form if
one chooses $\vk_{+}=-1$ and $\bar\vk_{-}=-1$. With this choice 
the flow equations with respect to $\b_\pm$ and $\bar\b_\pm$ are
\bal\la{vk1c}
\vk_+=-1\,, \quad\vk_-=1\,, \quad\bar\vk_+=1\,, \quad\bar\vk_-=-1\,,\quad \vk_{+-}=1\,,\quad \bar\vk_{+-}=1\,,
\eal
\bal\la{eqbep3}
{\pa\cH_{A}\ov\pa \b_{-}} &=2\cJ\bT_-\,,\qquad 
{\pa\cH_{A}\ov\pa \b_{+}} =2\cJ\bT_+\, +\, \b_{-}\bT\overline\bT\,,
\\
{\pa\cH_{A}\ov\pa \bar\b_{+}} &= 2\bar\cJ\bT_+
\,,\qquad
{\pa\cH_{A}\ov\pa \bar\b_{-}} = 2\bar\cJ\bT_-
\,+\, \bar\b_{+}\bT\overline\bT\,.
\eal
Calculating the coefficients $\cG_i$ with this choice of $\vk$'s for a deformed CFT which had  left- and right-moving conserved currents before the deformation, see appendix \ref{examples}, one finds that they are independent of $\b_+$ and $\bar\b_-$ (in fact with all the ten parameters switched on), and, therefore, the Hamiltonian density $\cH_A$ is independent of these two parameters too. Thus, the flow equations with respect to  $\b_+$ and $\bar\b_-$ lead to the existence of the following two relations
\bal\la{eqbep3c}
2\cJ\bT_+\, +\, \b_{-}\bT\overline\bT&=0\,,\\
2 \bar\cJ\bT_- + \bar\b_{+}\bT\overline\bT&=0\,,
\eal
in a deformed CFT. One might think that these relations imply the existence of improved left- and right-moving currents but it is so only in the case of a single $J\overline T$ or $\bar J T$ deformation.

%%%%%%%%%%%%%%%%%%%%%%%%%%%%%%%%%%%%
\subsection{Flow equations for the energy in CFT conventions}

To write the flow equations  for the energy  in CFT conventions we introduce the charges $\rQ_\pm$ and $\cQ_\pm$ of the  left- and right-moving currents \eqref{jpm} and \eqref{imjpm}
\bal\nn
{\rQ_\pm\ov R}=\bra n|\rJ_\pm^0|n\ket = -{1\ov 2R}(A_{11}P_1\pm {1\ov A_{11}}\widetilde P^1)\,,
\eal
and similar expressions for the improved charges $\cQ_\pm$ of the   currents  \eqref{imjpm}.
 Then, we get
\bal\nn
{\pa E_n\ov \pa\rQ_\pm}={\pa E_n\ov \pa\cQ_\pm}= -\big({1\ov A_{11}}{\pa E_n\ov \pa \bP_1}\pm  A_{11}{\pa E_n\ov \pa\widetilde \bP^1}\big)\,,
\eal
and therefore
\bal\nn
\bra n|\rJ_\pm^1|n\ket =\bra n|\cJ_\pm^1|n\ket = {1\ov 2}\big(A_{11}{\pa E_n\ov \pa\widetilde P^1} \pm {1\ov A_{11}}{\pa E_n\ov \pa P_1}  \big) = \mp {1\ov 2}{\pa E_n\ov \pa\cQ_\pm}\,.
\eal
We also need
\bal\nn
\bra n|\bT^1{}_\pm|n\ket ={1\ov 2}\big(\bra n|\bT^1{}_1|n\ket \pm \bra n|\bT^1{}_0|n\ket  \big) =
{1\ov 2}\big({\pa E_n\ov\pa R} \pm {1\ov R}(1\mp v){\pa E_n\ov\pa v}   \big)\,,
\eal
\bal\nn
\bra n|\bT^0{}_\pm|n\ket ={1\ov 2}\big(\bra n|\bT^0{}_1|n\ket \pm \bra n|\bT^0{}_0|n\ket  \big) =
{1\ov 2 R}\big(-\check P+{\cQ_+^2-\cQ_-^2\ov R} \pm  (1\mp v) E_n  \big)\,.
\eal

\medskip

By using these relations and eqs.(\ref{eqbep3}), one gets the following flow equations for the energy of the state $|n\ket$ (with $A_{--}=1$)
\bal\la{eqa11encft}
A_{11}{\pa E_n\ov\pa A_{11}} &= \cQ_+ {\pa E_n\ov\pa\cQ_-} +\cQ_- {\pa E_n\ov\pa \cQ_+}  \,,
\eal
\bal\la{eqapmencft}
{\pa E_n\ov\pa \a} &= -E_n{\pa E_n\ov \pa R} - {\check P-{\cQ_+^2-\cQ_-^2\ov R}\ov R}{\pa E_n\ov \pa v}\,,
\\
{\pa E_n\ov\pa \b_-} &={\cQ_+}\big({\pa E_n\ov \pa R}- {1\ov R}(1+ v){\pa E_n\ov\pa v}  \big)-{1\ov2}{\pa E_n\ov \pa\cQ_+}\big(\check P-{\cQ_+^2-\cQ_-^2\ov R}+(1+v) E_n\big)\,,
\\
{\pa E_n\ov\pa \bar\b_+} &={\cQ_-}\big({\pa E_n\ov \pa R}+ {1\ov R}(1- v){\pa E_n\ov\pa v}  \big)+{1\ov2}{\pa E_n\ov \pa\cQ_-}\big(\check P-{\cQ_+^2-\cQ_-^2\ov R}-(1-v) E_n\big)\,,
\\
{\pa E_n\ov\pa \b_+} &={\cQ_+}\big({\pa E_n\ov \pa R}+ {1\ov R}(1- v){\pa E_n\ov\pa v}  \big)-{1\ov2}{\pa E_n\ov \pa\cQ_+}\big(\check P-{\cQ_+^2-\cQ_-^2\ov R}-(1-v) E_n\big) +{\b_-}{\pa E_n\ov\pa \a} \,,
\\
{\pa E_n\ov\pa \bar\b_-} &={\cQ_-}\big({\pa E_n\ov \pa R}- {1\ov R}(1+ v){\pa E_n\ov\pa v}  \big)+{1\ov2}{\pa E_n\ov \pa\cQ_-}\big(\check P-{\cQ_+^2-\cQ_-^2\ov R}+(1+v) E_n\big) +{\bar\b_+}{\pa E_n\ov\pa \a} \,.
\eal
As was discussed in the previous subsection, in the case of a CFT with left- and right-moving conserved currents ${\pa E_n\ov\pa \b_+} ={\pa E_n\ov\pa \bar\b_-} =0$.
%%%%%%%%%%%%%%%%%%%%%%%%%%%%%%%%%%%%

%%%%%%%%%%%%%%%%%%%%%%%%%%%%%%%%
\subsection{Deformed CFT with left- and right-moving  currents}\la{spectrumCFT}

In this subsection we use the  defining equation \eqref{eqc2} for the Hamiltonian density to find a solution of the flow equations (\ref{eqapmencft}) for  a deformed CFT with left- and right-moving conserved currents. The undeformed Lagrangian is given by \eqref{L0}, and before deforming the model we shift and rescale $x^1$ as discussed in section \ref{fleqHdcft} to reduce \eqref{L0} to the canonical form \eqref{L00} with  the undeformed Hamiltonian density
\bal\la{H0cft}
\cH_0={1\ov2}p_1^2+ {1\ov2} (x'^1)^2 + {1\ov2}G^{kl}(x)p_k p_l + {1\ov2} x'^k x'^l\,G_{kl}(x) \equiv \cK\,.
\eal
We set $A_{11}=1$ because the effect of the $\tilde JJ$ deformation was discussed in section 
\ref{fleqen}, and the energy of a $\tilde JJ$ deformed model is related to the energy of the model with $A_{11}=1$  by eq.\eqref{eqa11ensol}. To simplify the notations we also set the parameters $v^1$, $u_1$ to zero. The dependence of these parameters can be easily restored by shifting $p_1$ and $x'^1$ in the Hamiltonian density, or by replacing the charges $\rQ_\pm$ with the improved charges $\cQ_\pm$ in the energy eigenvalues.

\medskip

Let us assume that $\cH_A$ and the coefficients $\cG_i$ of the defining equation \eqref{eqc2}  are operators, and let us rewrite  \eqref{eqc2} in the following form
\bal \la{eqcH}
\bra n|\, \cG_{2}\, \cH_A^2  -\cG_{1}\, \cH_A  + \cG_{0} \, |n\ket =0\, \,.
\eal 
Clearly, this equation does not really make sense, because the products of operators require careful definitions. In particular, $\cG_1$ given by \eqref{G1ex3} depends on $px'$, $p_1$ and $x'^1$ which do not commute with $\cH_A$. This can be cured because $\cG_2$ does not depend on any fields, and therefore,  shifting $\cH_A$ by a linear combination of $px'$, $p_1$ and $x'^1$, one can remove all field-dependent terms from $\cG_1$.
Performing this shift, we bring eq.\eqref{eqcH} to the form\footnote{This is also the simplest form to check \eqref{Scftv2}.}
\bal \nn
\cG_{2}\,\bra n| (\cH_A - \mS_A)^2 |n\ket   -\widetilde\cG_{1}\,\bra n|(\cH_A - \mS_A) |n\ket   + \bra n|\widetilde\cG_{0} \, |n\ket =0\, \,,
\eal 
where 
\bal \nn
\la{G2ex3bb}  \cG_{2}&= \a (1 - v^2) + {1\ov2} \b_-^2 (1 + v)^2 +  {1\ov2} \bar\b_+^2 (1 - v)^2 \,,
\\
\widetilde \cG_{1}&= 1 - v^2 \,,
\\
\widetilde \cG_{0}= \cK + &{1\ov \cG_2}\Big[
\frac{1}{2} px' \left(\bar\b_+^2(1-v)^2-\b_-^2 (1+v)^2\right)
-(1-v^2) \big(\b_- \rJ_+^0 (1+v) + \bar\b_+ \rJ_-^0 (1-v)\big)\\
&+ (px')^2 \big(\bar\b_+^2 \b_-^2-\a^2\big)
+2\rJ_+^0 px' \big(\a\b_- (1+v)+\bar\b_+^2 \b_- (1-v)\big)\\
&-2\rJ_-^0 px' \big(\a \bar\b_+(1-v)+\bar\b_+ \b_-^2 (1+v)\big)
- \big(\b_-\rJ_+^0 (1+v)+\bar\b_+ \rJ_-^0 (1-v)\big)^2\Big] \,,
\\
 \mS_{A}&=  {1\ov \cG_2}\Big(\frac{1}{2} px' \left(-2 \a v-\bar\b_+^2 (1-v)+\b_-^2 (1+v)\right)+\b_- \rJ_+^0 (1+v)+\bar\b_+\rJ_-^0 (1-v)\Big)\,.
\eal 
Now, to get an equation for the spectrum of the deformed CFT with left- and right-moving conserved currents we use the following relations
\bal\la{HAEn}
\bra n|\cH_A |n\ket = {E_n\ov R}\,,\quad \bra n|\rJ_\pm^0 |n\ket = {\rQ_\pm\ov R}\,,
\eal
\bal\la{pxpP}
\bra n|px' |n\ket = -{P\ov R}= -{\check{k} - \rQ_+^2+\rQ_-^2\ov R^2}\,,\quad \bra n|\cK |n\ket = {E_n^{(0)}\ov R}={\check{\cE}_n^{(0)} +\rQ_+^2+\rQ_-^2\ov R^2}\,.
\eal
Here, as was discussed in section  \ref{fleqen}, we 
split off the world-sheet momentum $P=k/R$ into its intrinsic part $\check P = \check{k}/R$, and the part due to the rotation and winding in the $x^1$ direction, and do the same with the eigenvalues $E_n^{(0)}=\cE_n^{(0)}/R$ of the Hamiltonian \eqref{H0cft}.  It is worthwhile to mention that only $\check{\cE}_n^{(0)}$ always coincides with the corresponding part of the spectrum of the undeformed model. As to $\rQ_\pm$ dependent part, it depends on the point of view on the deformation. One may say that the range of $x^1$ does not change with the deformation and then $\rQ_\pm$ do not change either, or one may use that the deformation was generated by the chain of transformations \eqref{transfA} which explicitly changed the range of $x^1$ and, therefore, $\rQ_\pm$. In any case $\rQ_\pm$ are the charges measured in the deformed theory.
Finally, we use the following replacement rule
\bal\nn
\bra n| K_1K_2 |n\ket \to \bra n| K_1 |n\ket\bra n| K_2|n\ket\,,
\eal
for products of any two operators in (\ref{HAEn}, \ref{pxpP}).  Let us also note that the relation for $\bra n|\cK |n\ket$ in \eqref{pxpP} is strictly speaking a replacement rule too because the eigenvectors of the deformed Hamiltonian 
$H_A$ are different from those of $H_0$. The only justification for these rules we have is that they lead to the deformed eigenvalues $E_n$ which satisfy the flow equations   (\ref{eqapmencft}).

\medskip

By using these rules we find that the spectrum of the deformed CFT with left- and right-moving conserved currents is given by
\bal \la{eqEsol}
&\cA \big({E_n\ov R} - \cS_E\big)^2 -\big({E_n\ov R} - \cS_E\big)  + \cC  =0 \,,\\
&{E_n\ov R} = \cS_E +{1-\sqrt{1-4\cA\cC}\ov 2\cA} \,.
\eal 
Here 
\bal 
\la{Acoeff}  \cA&= \a + {b_-^2\ov2}   +  {\bar b_+^2\ov2}  \,,\\
\cA\cC&= \cA\,{\cE_n^{(0)}\ov \cR^2} +
\frac{k}{2\cR^2} \left(b_-^2-\bar b_+^2\right)
-\frac{1}{\cR}\big(b_- \rQ_+ + \bar b_+ \rQ_-\big)+ \frac{k^2}{\cR^4}\big(\,{\bar b_+^2 b_-^2}-\a^2\big)\\
&
-2{k\,\rQ_+\ov \cR^3}b_- \big(\a +{\bar b_+^2} \big)
+2{k\,\rQ_-\ov \cR^3} \bar b_+\big(\a +{b_-^2} \big)
-\frac{1}{\cR^2} \big( b_-\rQ_+ +\bar b_+ \rQ_-\big)^2 \,,\\
 \cS_E&= {k\,v\ov \cR^2} + {1\ov \cA}\big(\frac{k}{2\cR^2}\big(\bar b_+^2 -b_-^2\big)+\frac{1}{\cR} b_- \rQ_+ +\frac{1}{\cR} \bar b_+\rQ_-\big)\,.
\eal 
where
\bal
b_- =\b_-\,\sqrt{1+v\ov 1-v} \,,\quad\bar b_+ = \bar\b_+\,\sqrt{1-v\ov 1+v}\,,\quad \cR=R\,\sqrt{1-v^2} \,.
\eal
We see that all the dependence of $v$ is absorbed in the rescaling of $\b_-$, $\bar\b_+$ and $R$, and a shift of the energy by $P v/(1-v^2)$. This is in agreement with the consideration in appendix \ref{examples}. 
Our solution (\ref{eqEsol}, \ref{Acoeff})
seems to agree with the one proposed in \cite{Mezei2019a}
at least if one switches off their parameters of the  deformation by $\rJ_\pm^1$, and parameters of the $\rJ \T$ and $\bar\rJ \overline{\T}$ deformations. We see in particular that the parameter dependence of our coefficient $\cA$ and their coefficient $A$ is similar. To check a precise agreement it 
would be necessary to match the conventions, and it also
might be necessary to perform an extra redefinition of the parameter of the \TTb\ deformation. We have not tried to do it.
Then, at $v=0$ our solution completely agrees with the one proposed in \cite{Kutasov2019, Hashimoto} for closely related single-trace deformations by using dual strings on deformed $AdS_3$ if one relates the deformation parameters used in \cite{Hashimoto} to ours as follows
\bal\la{compHK}
\lambda = -{\alpha+ \beta_-\bar\beta_+\over R^2}\,,\  
\epsilon_- =- {\beta_-\over \sqrt{2}\,R}\,,\ 
\epsilon_+ = {\bar\beta_+\over \sqrt{2}\,R}\,,\ 
q_R = \sqrt{2}\,Q_+ \,,\ 
q_L =  \sqrt{2}\,Q_-\,,\ n=k\,.
\eal
Note that to get the agreement we have to shift $\a$ by a term proportional to $\b_-\bar \b_+$. Such a shift  ruins the nice structure of the flow equations (\ref{eqbep3}), and introduces \TTb\ terms on the r.h.s. of these equations. It would be interesting to understand how flow equations can be derived from  dual strings on deformed $AdS_3$ without first having found the spectrum.  In appendix \ref{examplessp} we present the spectrum deformed by one or two of the operators \TTb\,, \JTb\,, and \JbT.

%%%%%%%%%%%%%%%%%%%%%%%%%%%%%%%%%%
\section{Conclusions}

In this paper the light-cone gauge approach  \cite{Sfondrini18b,SF19a} to the \TTb\  deformed two-dimensional models  has been extended  to  include 
the most general ten-parameter deformation by U(1) conserved currents, stress-energy tensor and their various quadratic combinations.
We have found an explicit ten-parameter deformed Hamiltonian for a rather general system of scalars with an arbitrary potential, and used it to derive the flow equations with respect to the parameters, and the spectrum of a deformed CFT with  left- and right-moving conserved currents.
There are many open questions and generalisations of the approach. We mention just a few of them.

\medskip

Clearly,  the light-cone gauge approach can also be  used to study multi-parameter deformations of the system of scalars, fermions and chiral bosons introduced in \cite{SF19a}. Since these deformations breaks Lorentz symmetry it is natural to analyse non-Lorentz invariant models such as the  nonlinear matrix Schr\"odinger model. 

\medskip

The system of flow equations for the spectrum of a deformed model is complicated, and admits many solutions.  We could not find its solution which would be completely determined by the spectrum of the undeformed model. It would be interesting to understand if such a solution exists for a generic model. 
Even in the case of a CFT but without left- and right-moving conserved currents it is unclear how to find a unique solution.
  
\medskip

If  a model is integrable then there is a generalisation of  the \TTb\ deformation to more general higher-spin deformations introduced in \cite{SZ16}. 
It is expected that  they can be described in the light-cone gauge approach by  coupling strings to W gravity, see \cite{Hull93} for a review. 
Higher-spin conserved currents can be also used to study \JTb\ type deformations which break Lorentz invariance of an undeformed model, see  \cite{Tateo19,Mezei2019b} for a recent discussion.  Clearly, as soon as the higher-spin deformations of \cite{SZ16} are understood in the light-cone gauge approach, one can consider multi-parameter deformations involving the higher-spin  currents.

%%%%%%%%%%%%%%%%%%%%%%%%%%%%%%%%%%
\section*{Acknowledgements}

I would like to thank Tristan McLoughlin for useful discussions.

%%%%%%%%%%%%%%%%%%%%%%%%%%%%%%%%%%
\appendix

%%%%%%%%%%%%%%%%%%%%%%%%%%%%%%%%%%
\section{Canonical transformation matrices}\la{canmatr}

The matrices $A_x$, $B_x$,  $A_p$ and $B_p$ appearing in the canonical transformation \eqref{cantransf} are
\bal\nn
A_x=\left(
\begin{array}{ccccccc}
 1 & -\frac{\tilde a_{+-}}{A_{--}} & \frac{c_{1-} \tilde a_{+-}-a_{+1}}{A_{11}} & 0 & -c_{1-} b_{+1} & -b_{+1} A_{11} & 0 \\
 0 & \frac{1}{A_{--}} & -\frac{c_{1-}}{A_{11}} & 0 & 0 & 0 & 0 \\
 0 & -\frac{a_{1-}}{A_{--}} & \frac{1+c_{1-} a_{1-}}{A_{11}} & 0 & 0 & 0 & 0 \\
 0 & 0 & 0 & 1 & 0 & 0 & 0 \\
 0 & -\frac{b_{--}}{c_{1-} A_{--}} & \frac{b_{--}}{A_{11}} & 0 & 1 & 0 & 0 \\
 0 & 0 & 0 & 0 & 0 & A_{11} & 0 \\
 0 & 0 & \frac{1}{A_{11}} & 0 & 0 & 0 & \frac{1}{A_{11}} \\
\end{array}
\right)\,,
\eal
\bal\nn
B_x=\left(
\begin{array}{ccccccc}
 a_{+1} b_{+1} & 0 & 0 & -1 &\tilde a_{+-}& \frac{a_{+1}-c_{1-} \tilde a_{+-}}{A_{11}} & 0 \\
 0 & 0 & 0 & 0 & -1 & \frac{c_{1-}}{A_{11}} & 0 \\
 -b_{+1} & 0 & 0 & 0 & a_{1-} & -\frac{1+c_{1-} a_{1-}}{A_{11}} & 0 \\
 1 & 0 & 0 & 0 & 0 & 0 & 0 \\
 0 & A_{--} & 0 & 0 & \frac{b_{--}}{c_{1-}} & -\frac{b_{--}}{A_{11}} & 0 \\
 0 & 0 & A_{11} & 0 & 0 & 0 & -A_{11} \\
 0 & 0 & 0 & 0 & 0 & 0 & 0 \\
\end{array}
\right)\,,
\eal
\bal\nn
A_p=\left(
\begin{array}{ccccccc}
 1 & a_{+-} & a_{+1} & 0 & 0 & b_{+1} & 0 \\
 0 & A_{--} & 0 & 0 & 0 & 0 & 0 \\
 0 & a_{1-} A_{11} & A_{11} & 0 & 0 & 0 & 0 \\
 0 & 0 & 0 & 1 & 0 & 0 & 0 \\
 0 & \frac{b_{--}}{c_{1-}} & 0 & 0 & 1 & 0 & 0 \\
 0 & -\frac{b_{--}}{A_{11}} & 0 & 0 & -\frac{c_{1-}}{A_{11}} & \frac{1}{A_{11}} & 0
   \\
 0 & -a_{1-}A_{11} & -A_{11} & 0 & 0 & 0 & A_{11} \\
\end{array}
\right)\,,
\eal
\bal\nn
B_p=\left(
\begin{array}{ccccccc}
 0 & 0 & 0 & 0 & 0 & 0 & 0 \\
 0 & 0 & 0 & 0 & 0 & 0 & 0 \\
 0 & \frac{b_{--}}{A_{11}} & 0 & 0 & \frac{c_{1-}}{A_{11}} & 0 & 0 \\
 0 & 0 & 0 & 0 & 0 & 0 & 0 \\
 0 & -c_{1-} a_{1-} & -c_{1-} & 0 & 0 & 0 & 0 \\
 0 & 0 & 0 & 0 & 0 & 0 & 0 \\
 0 & 0 & 0 & 0 & 0 & 0 & 0 \\
\end{array}
\right)\,,
\eal
\bal\nn
C_x=\left(
\begin{array}{c}
 0 \\
 0 \\
 0 \\
 0 \\
 0 \\
 -c_{1-} \\
 0 \\
\end{array}
\right)\,,\quad
C_p=\left(
\begin{array}{c}
 0 \\
 -c_{1-} a_{1-} \\
 -c_{1-} \\
 0 \\
 0 \\
 0 \\
 0 \\
\end{array}
\right)
\eal
where 
\bal\nn
\tilde a_{+-} ={a_{+-}-a_{+1} a_{1-}}\,.
\eal

%%%%%%%%%%%%%%%%%%%%%%%%%%%%%%%%%%%%
\section{Virasoro  constraints coefficients}\la{virc}

The coefficients $\tilde G^{rs}$ and so on appearing in the transformed Virasoro constraint $C_2$
\eqref{C2f}
are given by
\bal\nn
\tilde G^{rs}  &=  G^{ab} (A_p)^r{}_{\hat a}(A_p)^s{}_\hb +G_{ab} (B_x)^{\ha r}(B_x)^{\hb s}\,,
\\
\tilde G_{rs}  &=  G_{ab} (A_x)^\ha{}_r(A_x)^\hb{}_s +G^{ab} (B_p)_{r\ha}(B_p)_{s\hb}\,,
\\
\hat G^{r}{}_{s}  &=  G^{ab} (A_p)^r{}_\ha(B_p)_{s\hb} +G_{ab} (B_x)^{\ha r}(A_x)^\hb{}_{s}\,,
\\
\tilde G^{rk}  &=  G^{ak} (A_p)^r{}_\ha\,,\quad  \tilde G_{rk}= G_{ak} (A_x)^\ha{}_r\,,
\\
\check G^{r}{}_{k}  &=  G_{ak} (B_x)^{\ha r}\,,\quad \check G_{r}{}^{k}= G^{ak} (B_p)_{r\ha}\,.
\eal
Here the indices $a,b = +,-,1$ are in one-to-one correspondence with $\ha,\hb = \hat1,\hat2,\hat3$, and $r,s=\hat1,\ldots,\hat7$.

The solution \eqref{cxpr} to  the constraints equations $\tilde C_1=0$, $C_a^V =0$ and $C^1_\cU=0$, and the light-cone gauge conditions $x^+=\tau$, $p_-=1$, $v^-=0$  is given by
\bal\la{cxp2}
\cX'^{\hat2}&=x'^-=v^+p_+ -\widetilde{p x}'+{c_{1-}\ov A_{11}} (x'^1-v^1) \,,
\\
\cX'^{\hat4}&=\tilde X'_+=-p_+\,,
\\
\cX'^{\hat5}&=\tilde X'_-=\frac{b_{--} }{c_{1-}A_{--}}(v^+p_+-\widetilde{p x}') -A_{--}+({1\ov A_{--}}-1) {b_{--}\ov A_{11}}(x'^1-v^1)\,,
\\
\cX'^{\hat6}&=\tilde X'_1=-p_1 + u_1+{c_{1-}\ov A_{11}}\,,
\\
\cX'^{\hat7}&=\Upsilon'^1=-x'^1\,,
\eal
where
\bal\nn
\widetilde{p x}' = (p_1-u_1) (x'^1-v^1)+p_kx'^k\,.
\eal

The coefficients $\cG_i$ appearing in \eqref{eqc2}
are given by
\bal 
\nn
\cG_{2}&={1\ov2} \tilde G^{++} + {1\ov2}\vk^r \vk^s \tilde G_{rs}+\hat G^{+}{}_{s} \vk^s \,,
\\
 \cG_{1}\, &=\tilde G^{+s'} \cP_{s'} +{1\ov2} \vk^r \chi^s \tilde G_{rs}+\tilde G^{r'k} \cP_{r'} p_k +\vk^r x'^k \tilde G_{rk}\\
&+\hat G^{+}{}_{s}\chi^s +\hat G^{r'}{}_{s} \cP_{r'} \vk^s+\check G^{+}{}_{k} x'^k + \vk^r p_k \check G_{r}{}^{k} \,.
\\
 \cG_{0}\, &= {1\ov2}\tilde G^{r's'} \cP_{r'} \cP_{s'} +{1\ov2} \chi^r \chi^s \tilde G_{rs}+\tilde G^{r'k} \cP_{r'} p_k + \chi^r x'^k \tilde G_{rk}\\
&+\hat G^{r'}{}_{s} \cP_{r'} \chi^s +\check G^{r'}{}_{k} \cP_{r'} x'^k + \chi^r p_k \check G_{r}{}^{k}+\tilde\cH_x \,,
\eal 
where $k=2,\ldots,n$ are the indices of the remaining transversal coordinates and momenta, and $r',s'=\hat2,\ldots,\hat7$.

%%%%%%%%%%%%%%%%%%%%%%%%%%%%%%%%%%%%
\section{Examples of deformed Hamiltonians}\la{examples}
In this appendix we consider deformations of a model with the Hamiltonian density
\bal\la{CHo}
\cH_0= {1\ov2}G^{\mu\nu}(x)p_\mu p_\nu + {1\ov2} x'^\mu x'^\nu\,G_{\mu\nu}(x) +V(x)=\cK+V\,,
\eal
where $\cK$ is the density of the kinetic term, and  $G_{\mu\nu}$ and $V$ do not depend on $x^1$. The target space metric of the string sigma model is given by \eqref{GMNrel}.  Models of this type include also models with rotational symmetries where one should first use spherical coordinates to bring the model's Hamiltonian to the form \eqref{CHo}.
In all considered examples  we switch off the deformations by $J^1$, $\widetilde J^1$, $T^1{}_1$ and $\widetilde JJ$, that is we set the parameters $v^1$, $u_1$ to zero, and $A_{--}$, $A_{11}$ to one. The dependence of these parameters can be easily restored if desired. We  use the notations  \eqref{aTTb}
\bal\nn
\aTTb \equiv a_{+-}\,,\quad \aJtTt \equiv -a_{+1}\,,\quad \aJTs \equiv a_{1-}\,,\quad \aJTt \equiv b_{+1}\,,\quad \aJtTs \equiv b_{--}\,,\quad v\equiv -v^+\,.
\eal

%%%%%%%%%%%%%%%%%%%%%%%%%%%%%%%%%%%%%%%%%%%%%
\subsection*{Deformation by \TTb\  and $T^1{}_0$}
In this case the only nonvanishing parameters are $\a\equiv \aTTb$ and $v\equiv -v^+$. Calculating the coefficients $\cG_i$, one gets
\bal 
\la{G2ex1}  \cG_{2}&= \tilde \a (1 - v^2) \,,\quad \tilde\a=\alpha(1+\alpha  V)\,,
\\
 \cG_{1} &=(1 - v^2) (1 + 2 \a V) - 2v\,  \tilde \a\, px'  \,,\quad px'\equiv p_\mu x'^\mu\,,
\\
 \cG_{0} &= \cK + V (1 - v^2) - 
  \tilde \a (px')^2- v  (1 + 2 \a  V)px'   \,.
\eal 
Solving the defining equation \eqref{eqc2}, one gets the deformed Hamiltonian density
\bal
\cH_{\a,v}={1\ov 2\tilde\a}\left(1-\sqrt{1-{4\tilde\a \cK \ov 1-v^2}+\frac{4  \tilde\a ^2 (px')^2 }{\left(1-v^2\right)^2}}\,\right) +{V\ov 1+\a V} -\frac{v\,px' }{1-v^2}\,.
\eal
The 2-parameter deformed model is in fact related to the \TTb\  deformed one. Indeed, if one performs the following rescaling of the momenta, and the world-sheet  space coordinate $\s$
\bal\la{pst10}
p_\mu\to \sqrt{1-v^2}\,p_\mu\,,\quad \s\to{\s\ov \sqrt{1-v^2}}\,,
\eal
then the deformed action \eqref{S4} transforms as
\bal\la{SAttbv}
S_A\to S_A=\int_{0}^{R\sqrt{1-v^2}}\, {\rm d}\s{\rm d}\tau\, \big( p_\mu \dot{x}^\mu
\,-\,{1\ov \sqrt{1-v^2}}(\cH_{\a,0}-v\,px' ) \big)\,.
\eal
The last term in the formula is just proportional to the world-sheet momentum $P$, and, therefore, the energy $E_{\a,v}(R,P_{v})$ of a state with momentum $P_v$ in the 2-parameter deformed model on a circle of circumference $R$ is related to the energy $E_{\a,0}(R\sqrt{1-v^2}\,,P_{0})$ of the corresponding state with momentum $P_0$ in the \TTb\  deformed model on a circle of circumference $R\sqrt{1-v^2}$ as
\bal\la{Ettbv}
E_{\a,v}(R,P_{v})= {1\ov \sqrt{1-v^2}}(E_{\a,0}(R\sqrt{1-v^2}\,,P_{0})+v\,P_{0} )\,.
\eal
Due to the momentum quantisation $P_v$ and $P_0$ are related as $P_v = \sqrt{1-v^2}\,P_{0}$ which also follows from the transformation \eqref{pst10}.
Clearly, this simple relation exists only because of the Lorentz invariance of the \TTb\ model.

%%%%%%%%%%%%%%%%%%%%%%%%%%%%%%%%%%%%%%%%%%%%%
\subsection*{Deformation by \TTb, $JT_0$, $JT_1$, $\tilde JT_0$ and $\tilde JT_1$}
In this case we set the parameter $v$ of the $T^1{}_0$ to 0: $v=0$. Calculating the coefficients $\cG_i$, one gets
\bal 
\la{G2ex2}  \cG_{2}&=  \aTTb +{1\ov2} \aJTt^2 G_{11} +{1\ov2} \aJtTt^2 G^{11} +  (\aTTb^2 -  \aJtTt^2 \aJTt^2) V\,,
\\
 \cG_{1}\, &= 
1 + \aJTs \big(p_{1} + (\aJtTs - \aJTt G_{11}) px'\big) \\
 &- 
    \aJtTt (p_\mu G^{\mu 1} - \aJTt px' + \aJtTs G^{11} px') + \aJtTs x'^{1} - 
    \aJTt  G_{1\mu}x'^\mu\\
    &+4 V \big[\aTTb \big(1 + \aJtTt \aJTt px' + \aJTs (p_{1} + \aJtTs px') + \aJtTs x'^{1}\big) \\
 &+\aJtTt \aJTt \big(\aJTt (p_{1} + \aJtTs px') + \aJtTt (\aJTs px' + x'^{1})\big)\big] \,,
\\
 \cG_{0}\, &=   \cK  + 
 {1\ov2}px' \big[\aJTs^2 G_{11} px' + \aJtTs^2 G^{11} px' - 
    2 \aJTt (p_{1} + \aJtTs px') \\
    &\qquad\quad- 
    2 (-\aJtTs p_\mu G^{\mu 1} + \aJTs \aJtTt px' + \aTTb px' + \aJtTt x'^{1}) + 
    2 \aJTs G_{1\mu}x'^\mu\big]\\
    &\qquad\quad+ V \big[ 
    \big(1 + \aJTs (p_{1} + \aJtTs px') + \aJtTs x'^{1}\big)^2 \\
       &\qquad\quad- \big(\aJTs \aJtTt px' + \aTTb px' + 
       \aJTt (p_{1} + \aJtTs px') + \aJtTt x'^{1}\big)^2\big]\,.
\eal 
The Hamiltonian density $\cH_A$ is then found by solving the defining equation \eqref{eqc2}. The resulting Hamiltonian is very complicated, and it is unclear how one could find its spectrum even in the CFT limit where $V=0$. The situation becomes better if $G_{11}=1$ (or a constant), and $G_{1k}=0\,,\, k=2,\ldots,n$. Then, the model has left- and right-moving conserved currents, and one can redefine the deformation parameters to simplify the Hamiltonian. This is done in the next subsection.

%%%%%%%%%%%%%%%%%%%%%%%%%%%%%%%%%%%%%%%%%%%%%
\subsection*{Deformed CFT with  left- and right-moving  currents} \la{DefCFT}
In this subsection we use parameters $\a\equiv \aTTb$, $v\equiv -v^+$ and $\b_\pm$, $\bar\b_\pm$, and  consider deformations of a CFT with the Hamiltonian
\bal\la{CHocft}
\cH_0={1\ov2}p_1^2+ {1\ov2} (x'^1)^2 + {1\ov2}G^{kl}(x)p_k p_l + {1\ov2} x'^k x'^l\,G_{kl}(x) \equiv \cK\,,
\eal
 by \TTb, $J T_+ \leftrightarrow J\T$, $J T_- \leftrightarrow J\overline T$, 
$\bar J T_+ \leftrightarrow  \bar J T$, $\bar J T_- \leftrightarrow \bar J\overline \T$ and $T^1{}_0$.

Calculating the coefficients $\cG_i$ by choosing $\vk$'s as in  \eqref{vk1c}, one gets
\bal 
\la{G1ex3}  \cG_{2}&= \a (1 - v^2) + {1\ov2} \b_-^2 (1 + v)^2 +  {1\ov2} \bar\b_+^2 (1 - v)^2 \,,
\\
 \cG_{1}\, &= 
 1-v^2 - 
   2 v  \a px'  +\big(\b_-^2  (1 + v)- \bar\b_+^2(1 - v)\big)px'  \\
   &- \bar\b_+  (p_1 - x'^1) (1 - v) - 
   \b_-  (p_1 + x'^1)(1 + v)\,,
\\
 \cG_{0}\, &=   \cK  -  \a (px')^2
 + px' \big(
 \bar\b_+  (p_1 - x'^1)- \b_-  (p_1 + x'^1)+
  {1\ov2}(\bar\b_+^2 + 
   \b_-^2) px' -  v \big)\,.
\eal 
We see that $\cG_i$, and therefore the deformed Hamiltonian, do not depend on $\b_+$ and $\bar\b_-$. 
Thus, for a CFT with  left- and right-moving conserved currents there is no deformation by $J T_+ \leftrightarrow J\T$ and $\bar J T_- \leftrightarrow \bar J\overline \T$ if one chooses correctly the parameter of the \TTb\ deformation. Let us also mention that if the potential $V$ does not vanish then $\cG_i$ do depend on $\b_+$ and $\bar\b_-$. The Hamiltonian of the model is obviously given by
\bal
\cH_A =\cH(R,\a,\b_-,\bar\b_+,v)= {\cG_1 -\sqrt{\cG_1^2 - 4\cG_2\cG_0}\ov 2\cG_2}\,,
\eal
but its surprising feature is that its $v$-dependence is very similar to the one in \eqref{SAttbv}.
Indeed it is easy to check that the deformed action
\bal\la{Scftv1}
S_A=\int_{0}^{R}\, {\rm d}\s{\rm d}\tau\, \big( p_\mu \dot{x}^\mu
\,-\,\cH(\a,\b_-,\bar\b_+,v)\big)\,,
\eal
satisfies 
\bal\la{Scftv2}
S_A=\int_{0}^{R\sqrt{1-v^2}}\, {\rm d}\s{\rm d}\tau\, \big( p_\mu \dot{x}^\mu
\,-\,{1\ov \sqrt{1-v^2}}(\,\cH(\a,b_-,\bar b_+,0)-v\,px'\, ) \big)\,.
\eal
where 
\bal
b_- =\b_-\,\sqrt{1+v\ov 1-v} \,,\quad\bar b_+ = \bar\b_+\,\sqrt{1-v\ov 1+v} \,.
\eal
Thus, up to the rescaling of $\b_-$, $\bar\b_+$, the energy of the model with nonvanishing $v$ is related to its energy with $v=0$ as in
\eqref{Ettbv}.

\bigskip

Let us also calculate the coefficients $\cG_i$ with another interesting
 choice of $\vk$'s 
\bal\la{vk2c}
\vk_+=-1\,, \quad\vk_-=-1\,, \quad\bar\vk_+=-1\,, \quad\bar\vk_-=-1\,,\quad \vk_{+-}=1\,,\quad \bar\vk_{+-}=1\,.
\eal
Then, the flow equations with respect to $\b_-$ and $\bar\b_+$ take the form
\bal\la{eqbem3d}
{\pa\cH_{A}\ov\pa \b_{-}} =2\cJ\bT_-\, -\, {\b_{-} }\bT\overline\bT\,,\qquad
{\pa\cH_{A}\ov\pa \bar\b_{+}} = 2\bar\cJ\bT_+
- {\bar\b_{+}}\bT\overline\bT\,.
\eal
The coefficients $\cG_i$ with $\vk$'s as in  \eqref{vk2c} are given by
\bal 
\la{G1ex3b}  \cG_{2}&= \a (1 - v^2) +  \b_-^2 (1 + v)v -  \bar\b_+^2 (1 - v)v \,,
\\
 \cG_{1}\, &= 
 1-v^2 - 
   2 v  \a px'  +\big(\b_-^2  (1 + 2v)- \bar\b_+^2(1 - 2v)\big)px'  \\
   &-\bar\b_+  (p_1 - x'^1) (1 - v) - 
  \b_-  (p_1 + x'^1)(1 + v)\,,
\\
 \cG_{0}\, &=   \cK  -  \a (px')^2
 +  px' \big(
 \bar\b_+  (p_1 - x'^1)- \b_-  (p_1 + x'^1)+
 (\bar\b_+^2 + 
   \b_-^2) px' -  v \big)\,.
\eal 
Now, setting $v=0$, one gets
\bal 
\la{G2ex3c}  \cG_{2}&= \a \,,
\\
 \cG_{1}\, &=  1  + \big(\b_-^2 - \bar\b_+^2\big)px'  - \bar\b_+  (p_1 - x'^1)  - 
  \b_-  (p_1 + x'^1)\,,
\\
 \cG_{0}\, &=   \cK  -  \a (px')^2
 +  px' \big(
 \bar\b_+  (p_1 - x'^1)- \b_-  (p_1 + x'^1)+
  (\bar\b_+^2 + 
   \b_-^2) px' \big)\,.
\eal 
Finally, setting $\a=0$, and solving the defining equation  \eqref{eqc2}, one gets the deformed Hamiltonian 
\bal\la{Hcft2} 
\cH_{\b_-,\bar\b_+}={\cG_0\ov \cG_1} = { \cK  -  \a (px')^2
 +   px' \big(
 \bar\b_+  (p_1 - x'^1)- \b_-  (p_1 + x'^1)+
 (\bar\b_+^2 + \b_-^2) px' \big)\ov 1  +\big(\b_-^2 - \bar\b_+^2\big)px'  - \bar\b_+  (p_1 - x'^1) - \b_-  (p_1 + x'^1)}\,,
\eal
which satisfies the flow equations  (\ref{eqbem3d}).

The flow equations are now more complicated but
a curious feature of this choice is that setting $v=\a=0$, one gets Hamiltonian \eqref{Hcft2} which is a rational function of the coordinates and momenta. It suggests that 
it is the \TTb\ deformation which is responsible for the square root form of the deformed Hamiltonian.

%%%%%%%%%%%%%%%%%%%%%%%%%%%%%%%%%%%%
\section{Examples of the deformed spectrum}\la{examplessp}

In this appendix we specialise eqs.(\ref{eqEsol}, \ref{Acoeff}) 
to the one- and two-parameter deformations.

%%%%%%%%%%%%%%%%%%%%%%%%%%%%%%%%%%%%%%%%%%%%%
\subsection*{Deformation by \TTb}

Setting $v=0$, $\b_-=0$, $\bar \b_+=0$, one gets 
\bal 
\la{Acoeffex1}  
{E_n\ov R} &= {1-\sqrt{1-{4\a\ov R}E_n^{(0)} +\frac{4 \a^2}{R^2}P^2}\ov 2\a} \,,
\eal 
which is a well-known result.  

%%%%%%%%%%%%%%%%%%%%%%%%%%%%%%%%%%%%%%%%%%%%%
\subsection*{Deformation by \JTb\ and $\bar J T$}

Setting $v=0$, $\a=0$, $\bar \b_+=0$, one gets 
\bal 
\la{Acoeffex2} 
{E_n+P\ov R} &=\frac{2\rQ_+}{R\, \b_-} +{1-\sqrt{\big(1+{2\b_-\ov R} \rQ_+\big)^2 -{2\b_-^2\ov R} (E_n^{(0)}+P)}\ov \b_-^2} \,,
\eal 
which agrees with \cite{Kutasov2018}. 

\medskip

Let us also give the formula for the $\bar J T$ deformation
\bal
{E_n^{\bar J T}-P\ov R} &=\frac{2\rQ_-}{R\, \bar\b_+} +{1-\sqrt{\big(1+{2\bar\b_+\ov R} \rQ_-\big)^2 -{2\bar\b_+^2\ov R} (E_n^{(0)}-P)}\ov \bar\b_+^2} \,.
\eal 

%%%%%%%%%%%%%%%%%%%%%%%%%%%%%%%%%%%%%%%%%%%%%
\subsection*{Deformation by \TTb+\JTb}

Setting $v=0$,  $\bar \b_+=0$, one gets 
\bal 
\la{Acoeffex3} 
{E_n+P\ov R} &=\frac{2\a P+2\b_-\rQ_+}{(2\a+\b_-^2)R} +{1-\sqrt{\big(1+{2\a\ov R}P+{2\b_-\ov R} \rQ_+\big)^2 -{2(2\a+\b_-^2)\ov R} (E_n^{(0)}+P)}\ov 2\a+\b_-^2} \,,
\eal 
which agrees with \cite{Kutasov2019, Mezei2019a}. 

%%%%%%%%%%%%%%%%%%%%%%%%%%%%%%%%%%%%%%%%%%%%%
\subsection*{Deformation by \JTb+$\bar J T$}

Setting $v=0$,  $\a=0$, one gets 
\bal 
\la{Acoeffex4} 
&{E_n\ov R} =\frac{(\bar\b_+^2-\b_-^2)P+2\b_-\rQ_++2\bar\b_+\rQ_-}{(\b_-^2+\bar\b_+^2)R} \\
&+{1-\sqrt{\big(1+{2\b_- \rQ_++2\bar\b_+\rQ_-\ov R} \big)^2 -{2\b_-^2(E_n^{(0)}+P)+2\bar\b_+^2(E_n^{(0)}-P)\ov R}  -{4 \b_-\bar\b_+ P (\b_-\bar\b_+ P + 2 \b_- \rQ_-  - 2 \bar\b_+ \rQ_+ )\ov R^2}}\ov \b_-^2+\bar\b_+^2} \,.
\eal

%%%%%%%%%%%%%%%%%%%%%%%%%%%%%%%%%%%%
\section{How to check the flow equations}\la{flow-eq}

In this appendix we explain how one can check the general flow equations (\ref{eqvp}-\ref{eqa1m}) for the density of the Hamiltonian $\cH_A$, and the flow equations for the one-parameter deformations in section \ref{cantransform}. All the calculations have been done by using Mathematica. 

\medskip

Let us rewrite the defining equation \eqref{eqc2} in the following form
\bal \la{eqcHapp}
\cG_{2}\, \cH_A^2  -\cG_{1}\, \cH_A  + \cG_{0} =0\, \,.
\eal 
We assume that $\cG_2\neq 0$ which is a generic situation. Computing the differential of this equation, we get
\bal\la{DHA}
d\cH_A ={ \cH_A d\cG_{1} -\cH_A^2 d\cG_{2} - d\cG_{0} \ov 2\cG_{2}\, \cH_A  -\cG_{1}}={ \cH_A\big( d\cG_{1}- {\cG_{1} \ov\cG_{2} }d\cG_{2} \big)- d\cG_{0}+ {\cG_{0} \ov\cG_{2} }d\cG_{2} \ov 2\cG_{2}\, \cH_A  -\cG_{1}}\,,
\eal
where we have used \eqref{eqcHapp} to reduce the numerator to a linear function of $\cH_A$. 

\medskip

This formula  is used to find the stress-energy tensor \eqref{Tmunu}, the conserved current 
\eqref{Jmu},  the topological current \eqref{Jtmu}, and their various quadratic  combinations of the form $\e_{\a\b} K_1^\a K_2^\b$  as rational functions of $\cH_A$.
Writing the flow equations in the form
\bal\la{floweqapp}
{\pa\cH_A\ov \pa \a_{\cO_i}} - \sum_j c_{ij}\cO_j=0\,,
\eal
where $ \a_{\cO_i}$ are the parameters and $\cO_i$ are all the deforming operators, we see that the equation is a rational function with the denominator equal to $(2\cG_{2}\, \cH_A  -\cG_{1})^2$. Thus, multiplying a flow equation by the denominator, we get a polynomial in $\cH_A$ which can be reduced to a linear function in $\cH_A$ by using the defining equation  \eqref{eqcHapp}. This linear function must vanish if the flow equation holds. Finally, one computes the derivatives of the Virasoro constraint coefficients $\cG_i$ with respect to $\a_{\cO_i}$, $x'^1$ and $p_1$, and the contractions ${\pa \cG_i\ov \pa x'^\mu}x'^\mu$, $p_\mu{\pa \cG_i\ov \pa p_\mu}$, ${\pa \cG_i\ov \pa x'^\mu}{\pa \cG_k\ov \pa p_\mu}$, and checks that the linear functions indeed vanish.

\medskip

In the case of the ten-parameter deformation and the general target space metric $G_{MN}$ it takes several hours to check one flow equation. The computation is much faster if the undeformed model is Lorentz invariant, that is $G_{MN}$ is of the form
\eqref{GMNrel}, and  $G_{1k}=0$, $k=2,\ldots,n$.

\medskip

Let us finally mention that to fix the coefficients $c_{ij}$ in the flow equations \eqref{floweqapp} we have used the target space metric of the form $G_{+-}=1$,  $G_{11}\neq1$, $G_{ik}\neq0$, $i,k=2,\ldots,n$, and we have set all the other metric components  to 0. The solution we found is unique.

%%%%%%%%%%%%%%%%%%%%%%%%%%%%%%%%%%%%%%%

%%%%%%%%%%%%%%%%%%%%%%%%%%%%%%%%%%%%%%%%%%%%%%%%%%%%%%%%%%%%%%%%%%%%%%%%%%%%%%%%%%%%%%%%%%%%%%%%%%%%%%%%%%%%%%%%%%%%%%%%%%%%%%%%%%%%%%%%%%

\end{document}